\documentclass[%
 reprint,
nofootinbib,
 amsmath,amssymb,
 aps,
 prm,
]{revtex4-2}

\usepackage{graphicx,xcolor}
\usepackage{dcolumn}
\usepackage{bm}

\usepackage{siunitx}

\newcommand{\bfa}[1]{\mathbf{#1}} 

\begin{document}

\preprint{APS/123-QED}

\title{Isomorph Invariant Dynamic Mechanical Analysis:\\
A Molecular Dynamics Study}

\author{Kevin Moch}
\homepage{kevin.moch@udo.edu}
\affiliation{
Fakultät Physik, Technische Universität Dortmund, D-44221 Dortmund, Germany
}

\author{Nicholas P. Bailey}
\affiliation{
 Glass and Time, IMFUFA, Department of Science and Environment, Roskilde University, P.O. Box 260, 4000 Roskilde, Denmark 
}

\date{\today}

\begin{abstract}
We simulate dynamic mechanical analysis experiments for the Kob-Andersen binary Lennard-Jones system. For this, the SLLOD algorithm with time-dependent strain rates is applied to give a sinusoidally varying strain at different densities and temperatures. The starting point is a temperature scan at a fixed reference density. Isomorph theory predicts that for other densities corresponding temperatures can be identified at which the mechanical properties are unchanged when scaled appropriately. We determine the isomorphically equivalent temperatures by analysing how particle forces change upon scaling configurations to the new density. Loss moduli expressed in suitable reduced units are compared for isomorphic state points. When plotted against the unscaled temperatures, these reduced loss curves are observed to collapse indicating the validity of isomorph theory for dynamic mechanical analysis experiments. Two different methods to determine isomorphic temperatures are considered. While one of them breaks down for the largest density rescalings considered in this study, the other one is still applicable in this region. The decorrelation of force vectors upon rescaling is investigated as a possible origin of this effect. Our results demonstrate that the simplification of the phase diagram entailed by isomorph theory for a wide class of system is relevant also for the mechanical properties of glasses.
\end{abstract}

\maketitle


\section{Introduction and Overview}
\label{sec:introduction}

The mechanical properties of glasses, among them metallic glasses, have received enormous attention in recent decades, due to their potential in applications, and the theoretical challenge of understanding their complex behavior~\cite{Eckert2007,Liu2015}. Properties depend dramatically on temperature and composition~\cite{Hand2010}, as well as thermal or processing history and loading rates~\cite{LouzguineLuzgin2012}. A variable which has received perhaps less attention is pressure or alternatively density~\cite{Molnr2017}.  Studies of glass-forming liquids have shown that in many cases a more complete picture emerges when density dependence is considered, because many dynamical properties depend on density and temperature simply through a combination $\rho^\gamma / T$ where $\rho$ is the density, $T$ the temperature and $\gamma$ the so-called density scaling exponent~\cite{Tolle/others:1998,Alba-Simionesco/others:2004,Casalini/Roland:2004}. The latter is often taken as a material constant in experimental work, but in general can depend on density~\cite{Boehling2012,Sanz/Others:2019,Casalini/Ransom:2020}. A theory which rationalizes this behavior has been under development for 13 years and is known as isomorph theory~\cite{Bailey2008a,Bailey2008b,Schroder2009,Gnan2009,Schroder2011,Ingebrigtsen2012,Schroder2014,Dyre2014}, the word {\em isomorph} referring to the cures in the phase diagram along which structural and dynamical properties are invariant, for example $T\propto \rho^\gamma$. The theory defines a class of materials which have good isomorphs, known as R-simple systems, and describes how to identify such materials, and the location of the isomorphs in the phase diagram, at least in computer simulations.

A common technique for studying mechanical properties in the glassy state is dynamical mechanical analysis (DMA) which involves subjecting the system to a small sinusoidal deformation and measuring the mechanical response via the complex elastic modulus, comprised of the real (storage) and imaginary (loss) parts~\cite{Rosner/Samwer/Lunkenheimer:2004}. The measurement is typically done over a range of temperatures at a fixed frequency at fixed (atmospheric) pressure. From the temperature profile of the response it is possible to study transitions from solid-like to fluid-like behavior, including the glass transition, as well as to observe relaxation processes in the solid. DMA has also been used in recent experiments involving large strain amplitudes on colloids to probe the connection between plastic deformation, micro-structure and excess entropy ~\cite{Galloway2020}. The protocol is also straightforward to implement in molecular dynamics simulations and has been used to study, for example, the Johari-Goldstein $\beta$-process in glasses~\cite{Cohen2012,Yu/Samwer:2014,Yu/others:2015,Yu/Richert/Samwer:2017}. In this work we investigate the consequences of R-simplicity for DMA using computer simulations of a binary Lennard-Jones glass. In particular the theory accurately predicts how the response at one  density, temperature and frequency can be used to predict that at a different density, with correspondingly different temperature and frequency.

The formal condition for a system to be R-simple is that for $\lambda\in\mathbb{R}$~\cite{Schroder2014}
\begin{equation}
U(\bfa{R}_1)> U(\bfa{R}_2) \Rightarrow U(\lambda\bfa{R}_1)> U(\lambda\bfa{R}_2)\;,
\end{equation} 
which states that changing density preserves the energy-ordering of micro-states. From this a number of interesting consequences follow, in particular the existence of {\em isomorphs}. Isomorph theory predicts the invariance of the structure and dynamics (expressed in suitable reduced units) of R-simple systems along isomorphs in the phase diagram~\cite{Gnan2009,Schroder2014}. In practice R-simple systems are identified as systems with strong potential energy $U$ and virial $W$ fluctuation correlations, i.e.,\ systems for which 
\begin{equation}
R:=\frac{\langle \Delta W \Delta U \rangle}{\sqrt{\langle (\Delta W)^2 \rangle \langle (\Delta U)^2 \rangle}}\gtrsim 0.9\;.
\label{eq:defR}
\end{equation}
The fluctuations $\Delta U$ and $\Delta W$ denote the difference between instantaneous values and their thermodynamic averages, while the sharp brackets in Eq.~\eqref{eq:defR} indicate $NVT$ ensemble averages. The correlation coefficient $R$ depends on the state point, and R-simplicity is typically confined to the condensed part of the phase diagram (pressures above the triple point pressure~\cite{Bailey2008a,Bailey/Others:2013}). For R-simple systems with $R<1$, this scale invariance and hence the existence of isomorphs is approximate. Since R-simplicity is typically not obvious from the form of the potential energy it is also referred to as ``hidden scale invariance''~\cite{Dyre2014}. 


Isomorph invariance applies to quantities which have been expressed in appropriately non-dimensionalized form, referred to as putting them into ``reduced units". The relevant factors by which quantities are reduced are based on macroscopic thermodynamic variables, most importantly the number density $\rho=N/V$ and the temperature $T$. This system for defining reduced units was pioneered by Rosenfeld who used the hard sphere model as a reference system for studying the relation between transport coefficients and the excess entropy~\cite{Rosenfeld1977}. One scales any quantity of interest by appropriate powers of $\rho$, $k_\text{B}T$ and the (average) particle mass $m$. For example for the distance $r$ between two particles, the reduced quantity is $\tilde r\equiv \rho^{1/3}r$; this essentially means expressing all lengths in units of the average interparticle spacing. For an elastic modulus, whose dimensions are the same as energy density, one divides by $\rho k_\text{B}T$, while the factor for reducing time is $\rho^{-1/3} (m/k_\text{B} T)^{1/2}$ which has a physical interpretation as the time to traverse the mean inter-particle spacing at the thermal velocity.

Recently, isomorph theory has been extended towards non-equilibrium physics, in particular physical aging~\cite{Dyre:2018,Dyre:2020}. A key challenge for glass-forming systems is identifying isomorphs in regions of the phase diagram where the equilibrium is not accessible due to excessively long relaxation times. Indeed, equilibrium entropy--more specifically the {\em excess entropy}, after subtracting the ideal gas term--plays a key role in isomorph theory, in that isomorphs in the equilibrium phase diagram can be identified as curves of constant excess entropy, or configurational adiabats. This leads to practical methods for determining isomorphs based on $W,U$ flucutations~\cite{Gnan2009}, but it is not {\em a priori} clear to what extent such methods apply in out-of-equilibrium situations. 

Isomorph invariance of steady-state shear response of single-component and Kob-Andersen binary Lennard-Jones~\cite{Kob/Andersen:1994,Kob/Andersen:1995a,Kob/Andersen:1995b} (KABLJ) liquids has been studied using the SLLOD algorithm~\cite{Evans/Morriss:1984,Ladd:1984} together with the Lees-Edwards boundary conditions~\cite{Lees/Edwards:1972,Allen/Tildesley:1987} using the molecular dynamics (MD) simulation package RUMD~\cite{Bailey2017} along isomorphs in regions of the phase diagram where the systems could be equilibrated~\cite{Separdar2013}. The isomorph invariance of the SLLOD algorithm was shown analytically in that work; an outcome of the analysis is the requirement that for isomorph invariance of sheared systems, the shearing rate must be fixed in reduced units. More recently isomorph invariance in the steady state shearing of KABLJ glasses was studied~\cite{Jiang2019}. There, collapse of the stress fluctuation statistics validates isomorph theory for steady-state non-equilibrium situations also far from the equilibrium-accessible part of the phase diagram. 

In this work, we apply a harmonically varying strain
\begin{equation}\label{eq:harmonic_strain}
    \epsilon(t) = \epsilon_0\sin(\omega t)
\end{equation}
to KABLJ glasses at different densities $\rho\ge 1.3$ and temperatures $T$ and measure the stress response $\tau_{xy}\equiv\tau$, with $\omega$ the angular frequency of the shear oscillation and $\epsilon_0$ the shear amplitude. The reason for not considering densities lower than 1.3 (for example the standard density of $\rho=1.2$) is to ensure strong correlations, $R>0.9$. In the linear response regime, the stationary stress response is given by
\begin{equation}
\tau(t)=\tau_0\sin(\omega t+\delta)+\tau_\text{off}\;,
\label{eq:stressfit}
\end{equation} 
with $\tau_0$ the stress response amplitude, $\delta$ the phase shift relative to the shear input, and $\tau_\text{off}$ accounting for initial stresses of the sheared systems, a computational artefact originating from their small sizes.
In DMA experiments, the storage and loss moduli are given by
\begin{equation}
G':=\frac{\tau_0}{\epsilon_0}\cos(\delta)\;,\;G'':=\frac{\tau_0}{\epsilon_0}\sin(\delta)
\end{equation}
and the loss modulus typically has a temperature dependence involving a dominant $\alpha$-peak indicating structural relaxation. In some materials a shoulder or separate peak corresponding to the so-called $\beta$-process is also observed~\cite{Rosner/Samwer/Lunkenheimer:2004}. The position of the $\alpha$ peak appears at a temperature for which the relaxation time $\tau_{\alpha}$ matches the frequency of the shear oscillation
\begin{equation}
\tau_{\alpha}=\frac{1}{\omega}\;.
\end{equation}

Isomorph invariance predicts that along isomorphs, the reduced loss modulus
\begin{equation}
\widetilde{G}'':=\frac{G''}{T\rho}
\label{eq:reducedloss}
\end{equation}
stays constant. Note that from now on we set the Boltzmann constant $k_\text{B}=1$. Hence, temperature dependent reduced loss curves obtained at different densities collapse as long as isomorphic temperatures are compared. To determine isomorphic temperatures, we utilize Schrøder's ``force-method''~\cite{Schroder2021} for determining  isomorphs from individual configurations. It is based on the reduced forces
\begin{equation}
\widetilde{\bfa{F}}=\frac{\bfa{F}}{T\rho^{1/3}}
\end{equation}
and the assertion that two state points $(\rho_1,T_1)$ and $(\rho_2,T_2)$ are isomorphic if the Frobenius norm of $\widetilde{\bfa{F}}$ for a typical configuration in one state point is identical to that for the same configuration (after uniformly scaling) in the other state point. Given a configuration at a starting state point 1, this yields the formula
\begin{equation}\label{eq:force_method}
T_2=\frac{|\bfa{F}_2|}{|\bfa{F}_1|}\bigg(\frac{\rho_1}{\rho_2}\bigg)^{1/3}T_1\;.
\end{equation}
Here, $\bfa{F}_i$ is the 3N-vector containing all forces acting on the particles on a particular configuration sampled from state point $(\rho_i,T_i)$. We note that for systems whose interactions are described by an inverse power law (IPL) with an exponent $n$, namely $v(r)\propto 1/r^n$, exact isomorphs exist. The formula for the isomorphic temperature $T_2$ in this case would be simply $T_2 =  (\rho_2/\rho_1)^{n/3}T_1$.

In this work, we extract reduced loss moduli curves at different densities and isomorphic temperatures and investigate whether these curves do collapse as predicted by isomorph theory. The structure of the paper is as follows. In Section~\ref{sec:sim_results} we briefly describe our procedures for generating independent configurations, cooling down to the glassy state and applying the DMA method, after which we present DMA profiles at different densities and show how they can be scaled onto each other. We consider two different methods to generate isomorphic configurations. In Sec.~\ref{sec:discussion_conclusions} we briefly discuss some implications of our results and draw conclusions.
\section{\label{sec:sim_results}Simulations and Results}
In this section, details of the performed simulations are given and results are presented. For all simulations, the shifted force method~\cite{Toxvaerd2011} with a cutoff $r_\text{cut}=2$ is imposed onto KABLJ systems. Outputs are saved for every 128\textsuperscript{th} time step.
The time steps for DMA at each state point are chosen such that the reduced time-step
\begin{equation}
\delta\widetilde{t}:=\frac{\delta t}{T^{-1/2}\rho^{-1/3}}\;.
\end{equation}
is kept constant with $\delta t(\rho=1.3,\;T=0.2)=0.005$. Keeping the same reduced time step accounts for trivial changes of timescale and makes it easier to ensure consistent simulations at different points along an isomorph. When not using reduced units we use the unit system defined by the energy scale $\epsilon_{AA}$, the length scale $\sigma_{AA}$ and the common mass $m_A=m_B$, which we refer to as Lennard-Jones (LJ) units.
\subsection{Equilibration and Cooling}
\label{sec:cooling}
\begin{figure}
    \centering
    \includegraphics[width=0.5\textwidth]{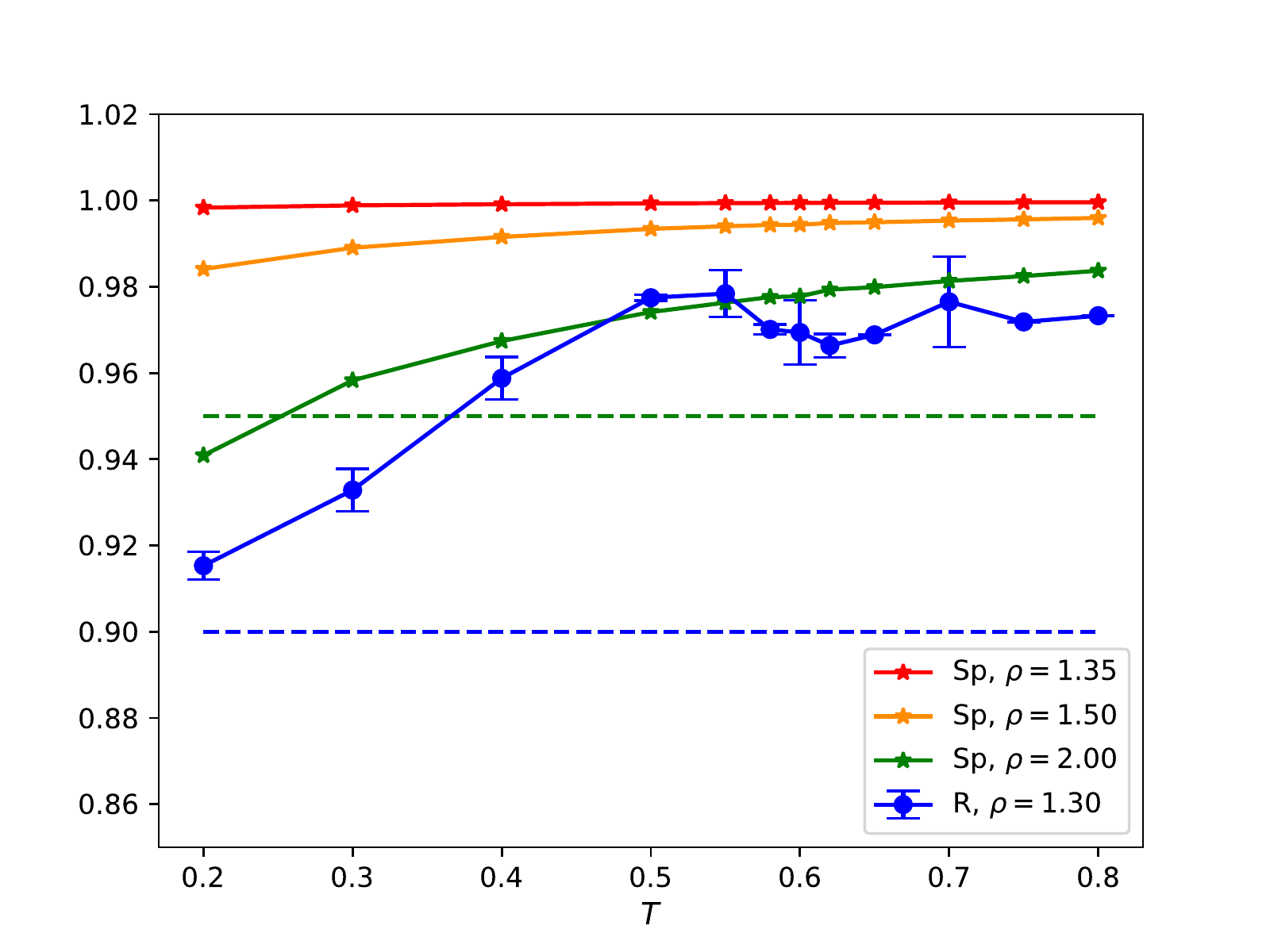}
    \caption{Pearson (blue) and Spearman (rest) correlation coefficients for different temperatures and densities.
    The Pearson $R$ is defined in Eq.~\eqref{eq:defR} and obtained at density $\rho=1.30$ during the DMA simulations, rather than equilibrium NVT simulations as normally required, and averaged over five independent runs. The dashed blue line marks the usual criterion for good isomoprhs, $R>0.9$. Error bars indicate the variance of the five different $R$-values obtained for the independent runs at each temperature. Spearman correlation values ("Sp") are of the force vectors before and after rescaling from $\rho=1.30$ configurations, as discussed at the end of section~\ref{sec:collapse}. Configurations are those drawn from the cooling run at the respective temperatures, i.e., those used for the temperature-matched method. The criterion proposed by Schr{\o}der~\cite{Schroder2021} for good isomorphs having Spearman correlations larger than 0.95 is indicated with the dashed green line.}
    \label{fig:Rplot}
 
\end{figure}

Five KABLJ systems, each having $N=8000$ particles, were initialized on a cubic lattice with density $\rho=1.3$ and particle identities randomly assigned. This lattice is unstable and therefore melts instantly when simulated at the temperature $T=1.2$. The melting temperature at density 1.2 is known~\cite{Pedersen/Schroder/Dyre:2018} to be 1.028; its value at density 1.3 can be roughly estimated as 1.55, thus the system at density 1.3 is supercooled, but crystallization to the equilibrium crystal phase cannot happen on the time scales we simulate. The systems are equilibrated by simulating $10^6$ time steps of size $\delta t=0.005$  using the NVT integrator (using a Nos\'e-Hoover-type thermostat). To check that the systems are equilibrated, the same simulation is repeated twice, each time using the final configuration from the previous simulation as the starting configuration, and the radial distribution functions as well as the mean-square displacements are compared. Since these are observed to be identical, we conclude that the system was properly equilibrated in the first run. These five independently equilibrated configurations were cooled down to $T=0.2$ at a cooling rate $1.6\times 10^{-5}$ in LJ units with time step $\delta t=0.005$. Data from DMA-production runs were averaged over the independent realizations. To check that the conditions for good isomorph invariance were satisfied we calculated the correlation coefficient $R$, Eq.~(\ref{eq:defR}), from the DMA runs. Since these are non-equilibrium simulations, this is not strictly the correct definition of $R$, nevertheless it can be used to get an idea of how well-correlated the pressure and energy fluctuations are. As can be seen in Fig.~\ref{fig:Rplot}, the values obtained lie between 0.91 and 0.98 thus satisfying the criterion for expecting good isomorphs. Moreover the pressure remains positive, even high (over 11 in LJ units), which for Lennard-Jones systems is sufficient to expect strong $W,U$ correlations and good isomorphs~\cite{Bailey2008a}.
Our starting point for shear simulations is given by the glass configurations generated by the procedure described here. This is in keeping with typical experimental protocols for  
DMA, whereby a metallic glass is made by cooling to for example room temperature, and subsequent mechanical testing at higher temperatures involves heating the glass~\cite{Rosner/Samwer/Lunkenheimer:2004,Yu/others:2015}.

\subsection{Generating Loss Curves}
\label{sec:losscurves}
To implement shear-oscillations, the SLLOD equations of motion are simulated, where the shear rate at each time step is set to the time-dependent value (following Eq.\eqref{eq:harmonic_strain})
\begin{equation}
\dot{\epsilon}(t)=\omega\epsilon_0\cos(\omega t)\;.
\end{equation}

\begin{figure}
    \centering
    \includegraphics[width=0.5\textwidth]{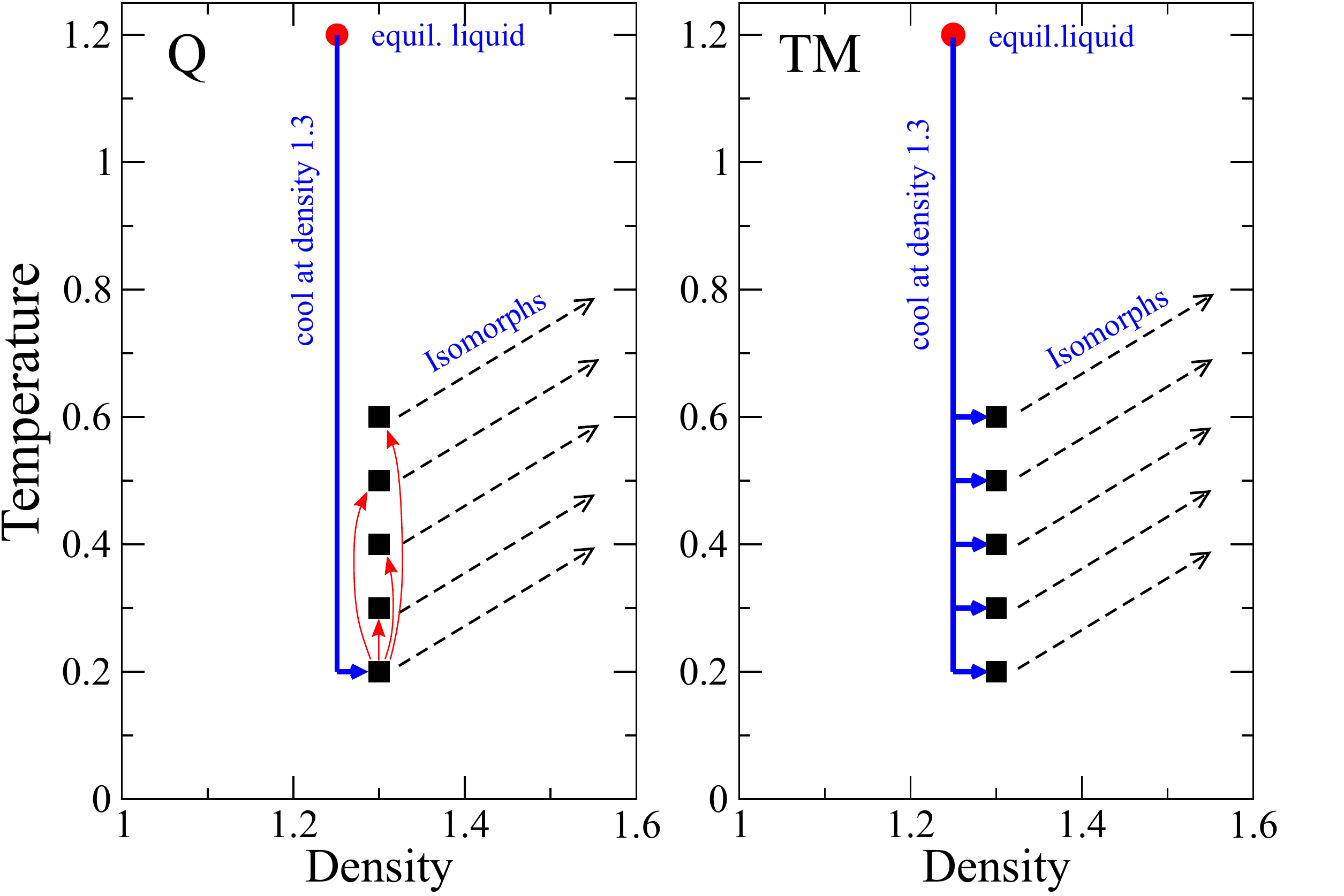}
    \caption{Schematic illustration of the two different protocols, or methods, we employ for the starting configurations for both the simulation runs and for determining isomorphic temperatures. The Q method illustrated on the left involves cooling at density 1.3 down to $T=0.2$ and using the resulting configurations for all subsequent simulations, scaling density and velocities as necessary. The TM method illustrated on the right involves cooling at density 1.3 to the relevant temperature for each DMA run at that density, and using those configurations both for the DMA runs and to determine isomorphic temperatures at high densities.}
    \label{fig:my_label}
\end{figure}

\begin{figure}
    \centering
    \includegraphics[width=0.5\textwidth]{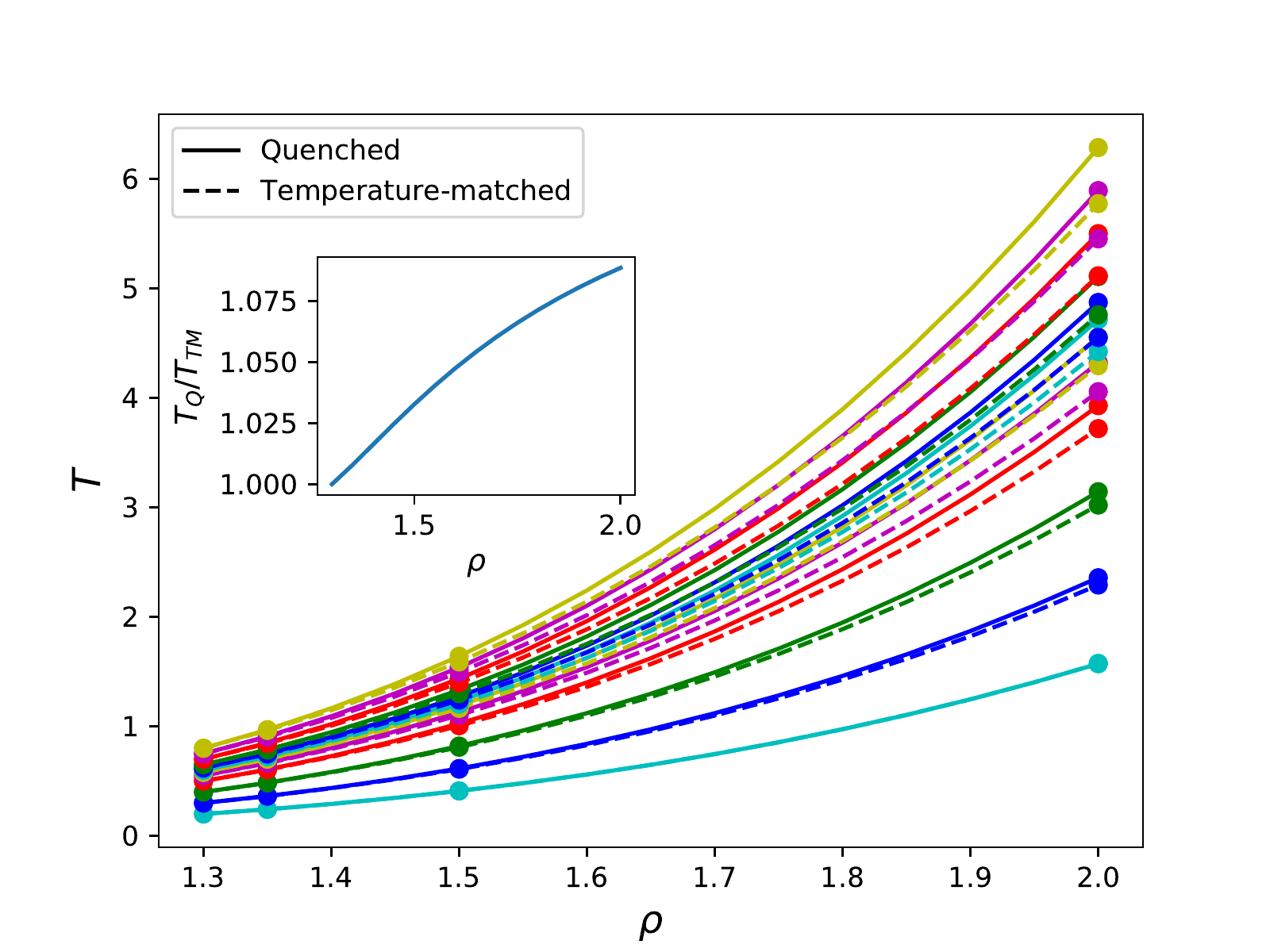}
    \caption{Isomorphs generated by the Q- (solid lines) and TM-method (dashed lines). The latter generates lower isomorph temperatures. Filled circles mark the state points at which the shear moduli presented in this work are obtained. The inset shows the ratio of isomorphic temperatures generated with the Q-method (``$T_Q$") to the analogous temperatures generated within the TM-method (``$T_{TM}$") for the isomorph starting at $T=0.6$ for $\rho=1.3$, highlighting again that the TM-method generates lower isomorphic temperatures.}
    \label{fig:isomorphs_schema}
\end{figure}

For the reference density $\rho_\text{ref}=1.3$ the angular frequency is chosen to be $\omega_\text{ref}=10^{-4}$ (in LJ units), and the simulations are carried out over a range of temperatures $T_{\text{ref},n}$, spanning the interval $[0.2,0.8]$ (indexed by $n$). The number of temperatures, $N_{\text{temp}}$, is chosen to give sufficient resolution in the DMA-profile without being too computationally demanding. For other, higher densities, the corresponding isomorphic temperatures are chosen according to the force method as explained above, yielding a family of isomorphs illustrated in Fig.~\ref{fig:isomorphs_schema}. As well as scaling the temperature, the angular frequency must be adjusted to keep its reduced-unit value constant. The time scale~\cite{Gnan2009} for this is $t_0(\rho, T) = \rho^{-1/3}\left(T/m\right)^{-1/2}$, as mentioned in the introduction. Therefore the reduced unit version of $\omega$ is $\tilde\omega\equiv t_0\omega$. Given that we choose the same reference frequency $\omega_\text{ref}$ for all temperatures at the initial density 1.3, the angular frequency at scaled densities is given by

\begin{equation}\label{eq:omega_rho}
\omega(\rho){=}\frac{\tilde\omega}{t_0(\rho,T)}
=\frac{\tilde{\omega}_\text{ref}}{(\rho_\text{ref}/\rho)^{1/3}(T_{\text{ref},n}/T_{\text{Iso},n}(\rho))^{1/2}}\;,
\end{equation}
with $T_{\text{Iso},n}(\rho)$ denoting the temperature at density $\rho$ which is isomorphic to $T_{\text{ref},n}$ at $\rho_\text{ref}=1.3$. Note that this expression suggests that at densities other than the reference density the angular frequency could depend on $n$, i.e., on the temperature, which would make it different from the standard DMA method, because the fixed-frequency temperature scan at one density would map to a non-fixed frequency temperature scan at a different density. When using the same configurations to initialize the DMA simulations, this ratio is independent of temperature, however, since the temperature ratio that appears in Eq.~\eqref{eq:omega_rho} depends only on the starting configuration and the two densities involved, see Eq.~\eqref{eq:force_method}.

In this study we pursued two different ways to generate isomorphic loss curves. In the first method we rescaled at all state points the same five quenched configurations obtained at the end of the cooling run of sec~\ref{sec:cooling} at $T=0.2$. Thus for temperatures apart the lowest (at a given density), each simulation was started with a configuration which was ``too cold", but was instantaneously heated to the desired temperature by rescaling velocities (our SLLOD algorithm uses an iso-kinetic thermostat~\cite{Evans/Morriss:2008} which conserves the kinetic energy). For the second method, we rescaled different configurations obtained at intermediate temperatures of the cooling run. These configurations are chosen such that the thermostat temperature at the corresponding point in the cooling run matched the temperature at which the DMA-run was to be carried out at the reference density. These configurations are presumably a better starting point for applying the force method as they are sampled from more or less the correct temperature. We label the former method as using quenched configurations, or the Q-method for brevity and the latter one as using temperature-matched configurations, or the TM-method. In Fig.~\ref{fig:my_label} these two different methods to obtain initial configurations is illustrated. The isomorphs generated by both methods are shown in Fig.~\ref{fig:isomorphs_schema}, where it can be seen that the temperature-matched configurations yield slightly lower isomorphic temperatures than the quenched configurations, most evident at the highest density 2.0. The ratio between isomorphic temperature and reference temperature is no longer independent of the latter for a given density change, so the adjusted frequency $\omega$ can now vary slightly for a given density, according to Eq.~(\ref{eq:omega_rho}). This turns out to be $\approx \SI{5}{\percent}$ at density 2.0. For all other studied densities, this change is even smaller.

\begin{figure}[h]
\centering
      \includegraphics[width=0.5\textwidth]{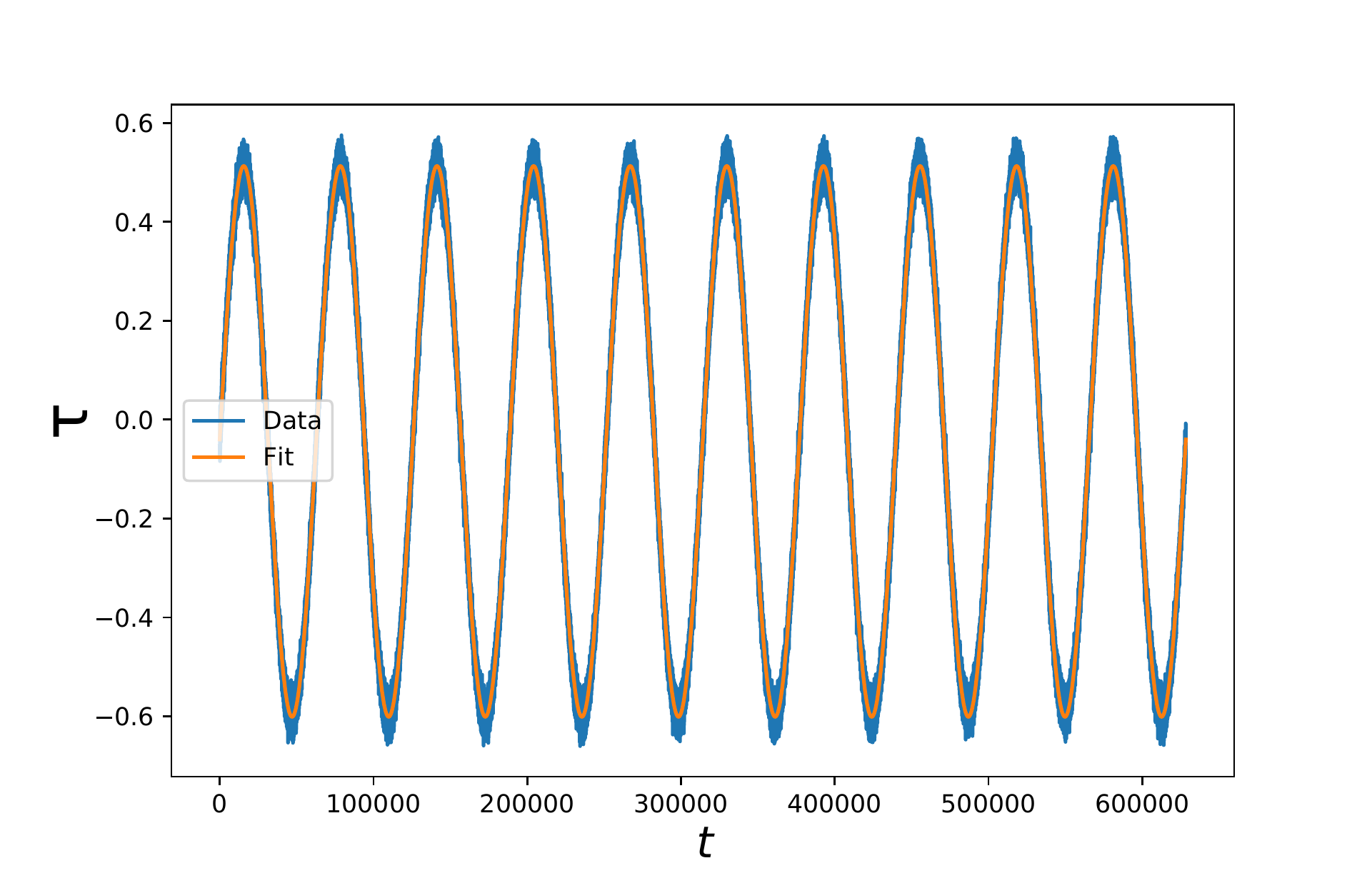}
      \caption{Stress response for $T=0.2$ and $\rho=1.3$ obtained for the angular frequency $\omega=10^{-4}$ (in LJ units) and shear amplitude $\epsilon_0=\SI{2}{\percent}$, averaged over five independently quenched configurations. The data points have been fitted using Eq.~\eqref{eq:stressfit}.}
      \label{fig:stress}
\end{figure}

For each state point 10 periods with shear amplitude $\epsilon_0=\SI{2}{\percent}$ were simulated. This shearing was applied to all five configurations of Sec.~\ref{sec:cooling} obtained during cooling after rescaling the volume and kinetic energy to study state points $(\rho,T)$. The measured stress responses are averaged over the five independent runs for each state point. In Fig.~\ref{fig:stress}, we show an example for the measured shear response at one state point, averaged over five independent runs.

%
\subsection{Collapse of Reduced Loss Curves}
\label{sec:collapse}
\begin{figure}
\centering
      \includegraphics[width=0.5\textwidth]{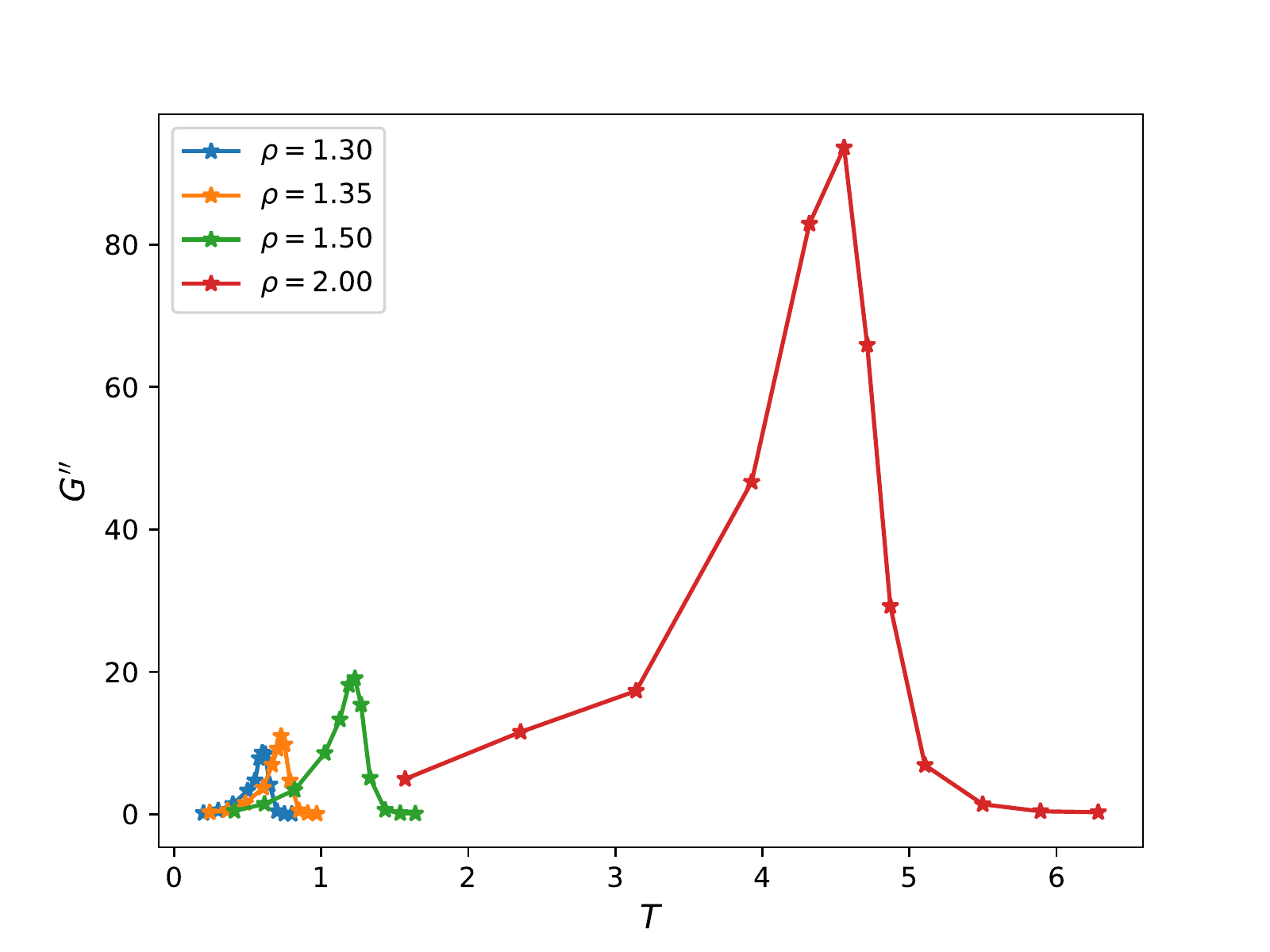}
      \caption{Loss curves in non-reduced form for four different densities, based on scaling initial configurations from the lowest density $\rho=1.3$ and using Eq.~\eqref{eq:force_method} to determine the appropriate temperatures at higher densities. Both the amplitude of the peak and its position on the temperature axis increase substantially with increasing density. These data are from the quenched method; that is the configurations were drawn from the lowest temperature of the cooling run ($T=0.2$) at the initial density ($\rho=1.3$) and the force method was applied to these configurations. Statistical errors were determined by fitting the individual runs and analyzing the run-to-run variation. These are smaller than the symbol sizes and hence not shown.}
      \label{fig:notreduced}
\end{figure}

\begin{figure}
\centering
      \includegraphics[width=0.5\textwidth]{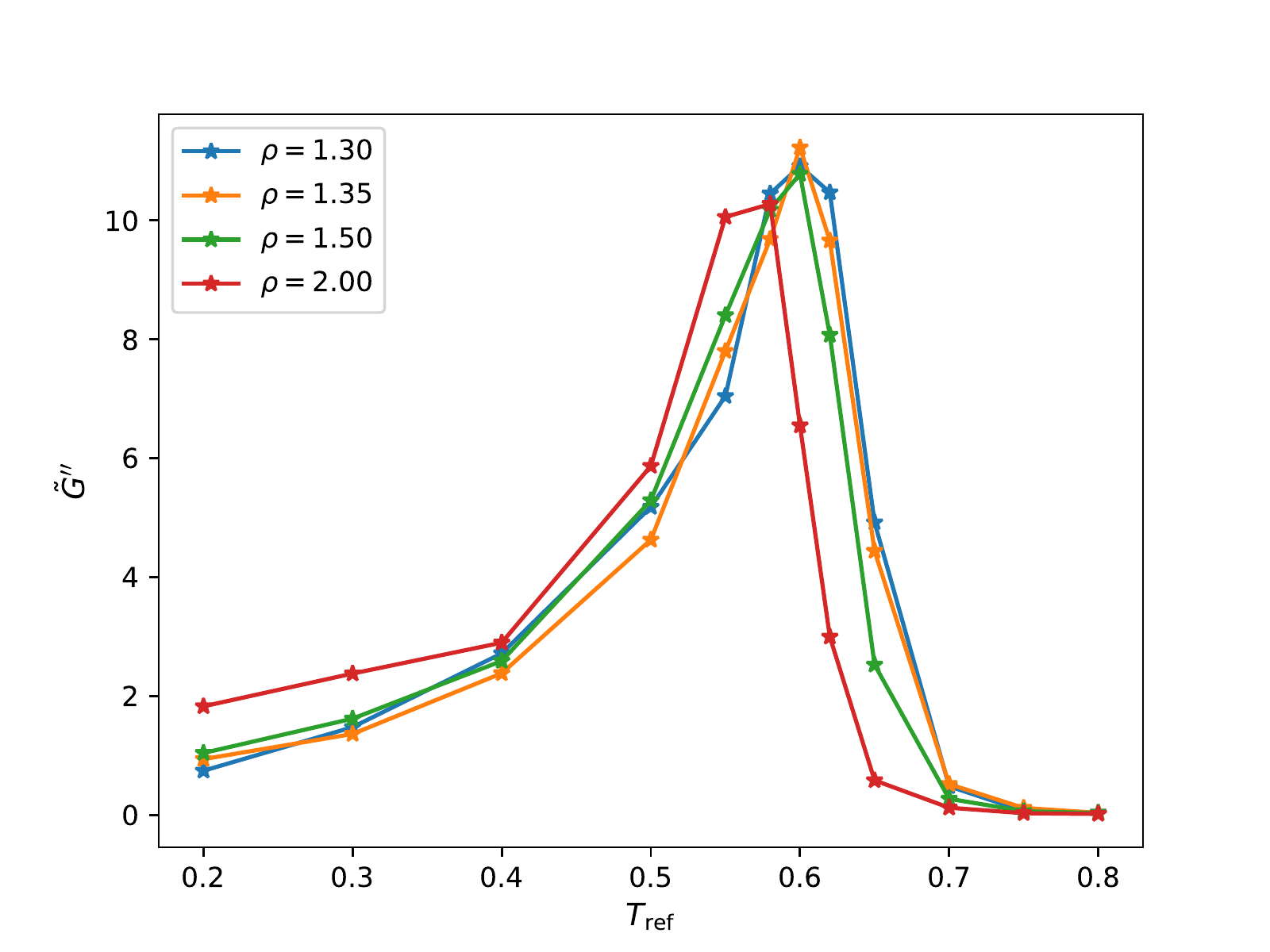}
      \caption{Same data as Fig.~\ref{fig:notreduced}, but the loss moduli have been put into reduced form according to Eq.~\eqref{eq:reducedloss}, and the x-axis now shows the reference temperature $T_\text{ref}$ instead of the actual temperatures $T$ used in the simulations (these are of course identical for density 1.3). Stars mark simulated points. Plotted this way, the three loss curves for densities up to $\rho=1.50$ within our final simulations show a good collapse. Going to larger densities $\rho=2.0$ (red), we see that the reduced loss curves do not collapse perfectly anymore. Note that the Q-method utilizes configurations cooled down to $T=0.2$ at density 1.3.}
      \label{fig:reduced}
\end{figure}

\begin{figure}
\centering
      \includegraphics[width=0.5\textwidth]{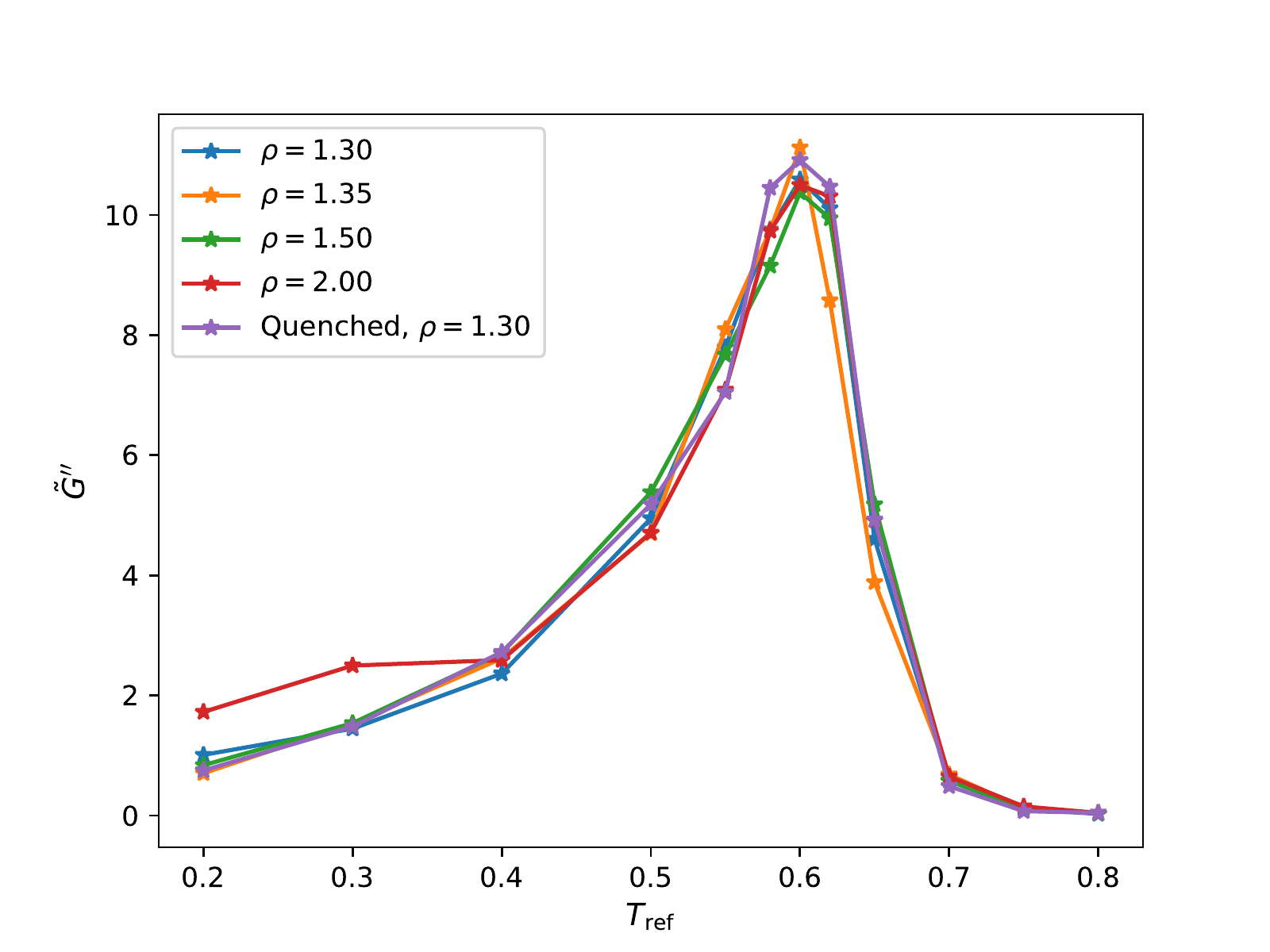}
      \caption{Analogous to Fig.~\ref{fig:reduced} but where the simulations have now been repeated after re-determining the isomorphic temperatures with the ``temperature-matched" method--using configurations drawn from the cooling run close to the temperature at which the deformation simulations were run (at the reference density). Again, stars mark simulated points. Here we observe a collapse of loss curves up to $\rho=2.0$ for which the Q-method lead to deviations, as demonstrated in Fig.~\ref{fig:reduced}. Only at the two lowest temperatures, the $\rho=2.0$ curve does not collapse onto the other curves. This can be understood from the Spearman correlations of Fig.~\ref{fig:Rplot} discussed at the end of section~\ref{sec:collapse}.}
      \label{fig:intermediate}
\end{figure}

In Fig.~\ref{fig:notreduced} we plot loss curves obtained with the Q-method for $\rho=1.3$, as well as $\rho=1.35$, $\rho=1.50$ and $\rho=2.00$ for corresponding isomorphic temperatures. We see that these four curves change substantially upon the increase of density, with both the peak temperature and the amplitude of the peak increasing strongly. However, the shapes are clearly similar, in itself suggesting an equivalence, and the possibility of a scaling collapse. As explained above, isomorph theory predicts in this case a collapse without any adjustable parameters. For this the moduli should be expressed in reduced units, Eq.~\eqref{eq:reducedloss}, and rather than the actual temperature, we use the reference temperature $T_\text{ref}$ on the x-axis. The result of this is shown in Fig.~\ref{fig:reduced} where we find that the three loss curves for $\rho=1.3, 1.35$ and $1.50$ now collapse, as predicted by isomorph theory. For even larger densities like $\rho=2.0$ the Q-method yields reduced loss curves which are not isomorph invariant. Here the force method, as applied to the quenched configurations, is no longer suitable for estimating the isomorph temperatures. On the other hand, using the TM-method, we still find a good agreement at $\rho=2.0$ as seen in Fig.~\ref{fig:intermediate}. Interestingly, for the lowest two values $T_\text{ref}$ the $\rho=2.0$ data fail to collapse with the other densities by a significant amount (the loss modulus in reduced units is nearly a factor of two larger than for the other densities). The TM-method cannot help here, as the difference is minimal (for the lowest reference temperature the two methods are the same since they use the same configurations). This can be understood considering the Spearman correlation values of the force vectors before and after rescaling presented in Fig.~\ref{fig:Rplot}. We discuss this in more detail below.

From Fig.~\ref{fig:isomorphs_schema} we see that applying the force method to quenched configurations leads to predictions for isomorphic temperatures which are larger than those from the temperature-matched configurations, which are expected to be more correct. Since in DMA experiments the loss modulus grows until it reaches a peak and then becomes smaller when going from small to large temperatures, this peak is reached earlier when the applied temperature has an positive offset. Hence the observed peak shift for $\rho=2.0$ to the left is consistent with the Q-method yielding temperatures larger than the ``true'' isomorphic temperatures. 

\begin{figure}
    \centering
    \includegraphics[width=0.5\textwidth]{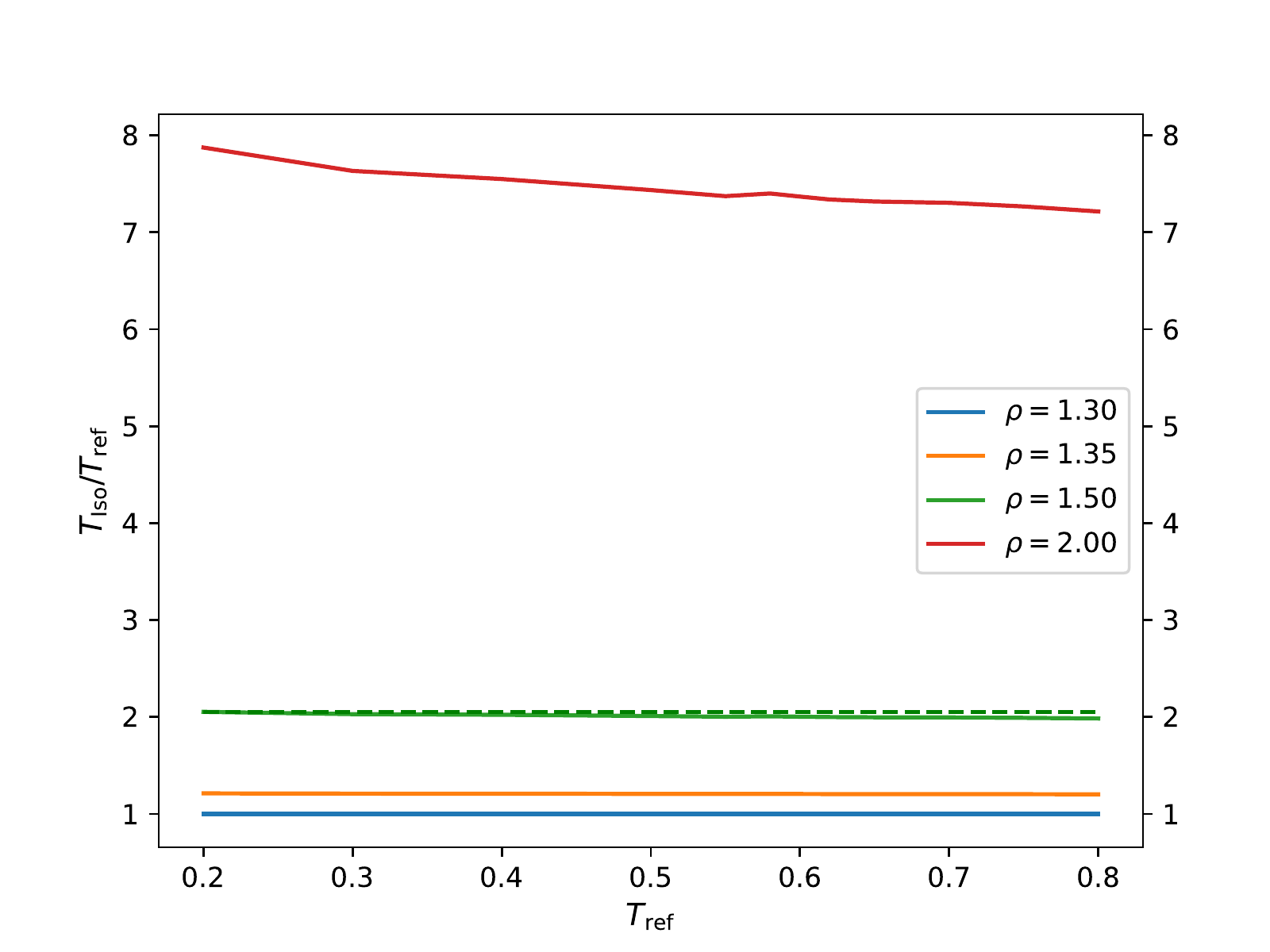}
    \caption{Temperature ratios between isomorphic temperatures obtained from the force method for different final densities and respective reference temperatures at density 1.3 as a function of these reference temperatures (that is, for configurations drawn from the cooling run when the temperature was the given $T_{ref}$). We observe a smaller ratio for configurations associated with higher temperatures. To show that some variation can be seen also $\rho=1.50$, we include a dashed horizontal line as a guide for the eye.}
    \label{fig:T_ratio_intermediate}
\end{figure}

In the following we explore why the TM-method yields better isomorphs. A useful diagnostic is the temperature ratio $T_2/T_1$ from Eq.~\eqref{eq:force_method} plotted for different configurations from the cooling run. These are shown in Fig.~\ref{fig:T_ratio_intermediate} for the different final densities. It shows, as expected, lower values of the temperature ratio for configurations drawn from higher temperatures, most evident at the highest density, $\rho=2.0$. The variation is nearly linear with $T_\text{ref}$. This is the root of the estimated isomorph temperatures being lower for the TM method than the Q-method (which corresponds to using the left-most point for all temperatures for each density). A simple argument for why this ratio is lower within the TM-method is that at higher temperatures there are more close encounters which effectively sample the LJ pair potential at shorter distances. The effective IPL exponent and therefore density scaling exponent $\gamma$ are known to decrease with decreasing interaction distance~\cite{Bailey2008b}, and the exponent determines how quickly the temperature rises along an isomorph. A lower density scaling exponent leads therefore to lower temperatures. 

\begin{figure}
\centering
      \includegraphics[width=0.5\textwidth]{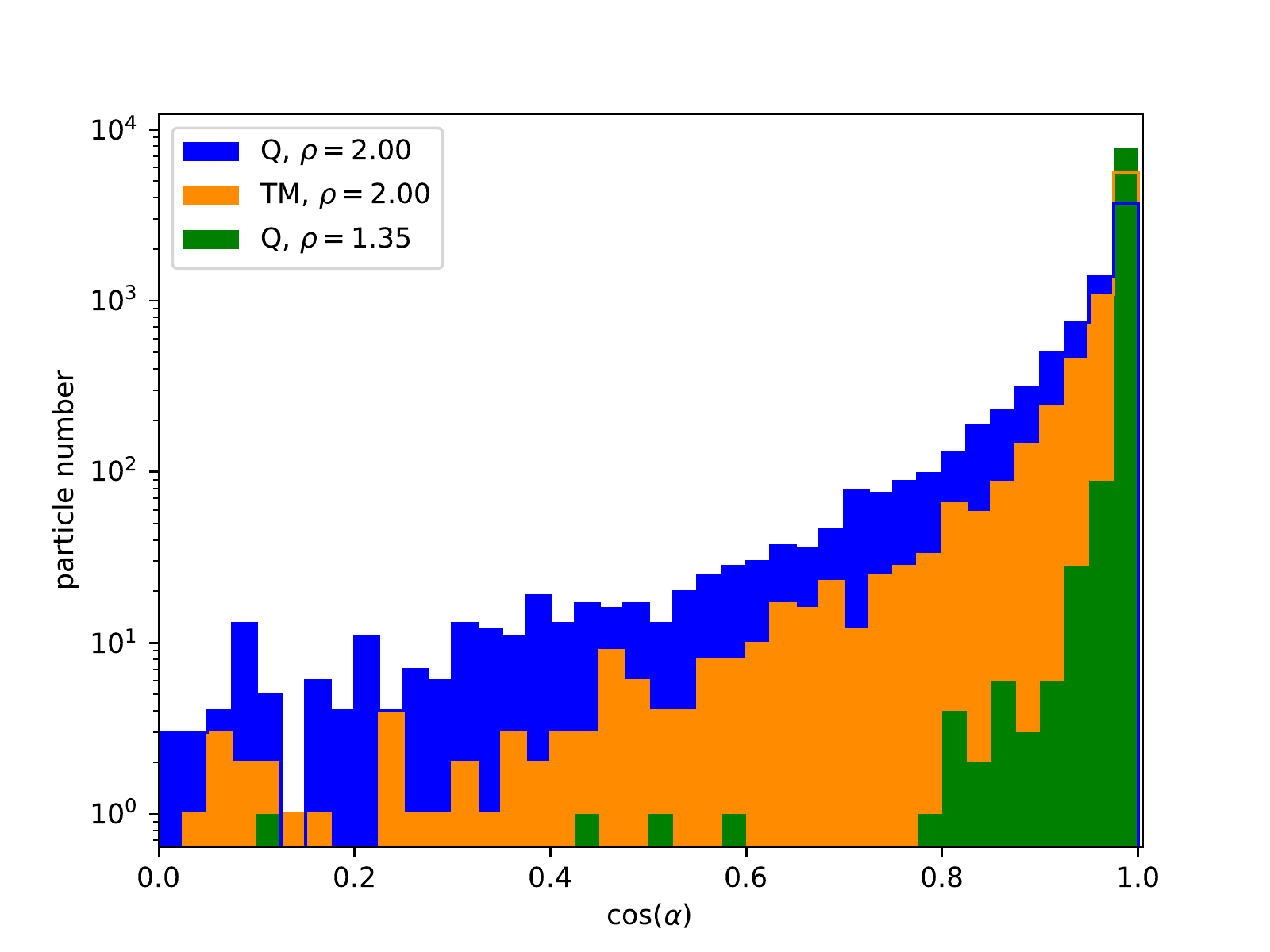}
      \caption{Histogram of the cosine of the angle $\alpha$ between forces before and after rescaling systems. The temperatures of the different systems in this plot are isomorphic to $T=0.6$ at $\rho=1.3$, cf. Fig.~\ref{fig:isomorphs_schema}. We observe that for larger densities, force directions change more when rescaling the systems. For $\rho=2.0$, the temperature-matched method (red) leads to a distribution of angles $\alpha$ concentrated more towards zero degrees ($\cos(\alpha)=1$) than the quenched (blue) method.}
      \label{fig:forceangles}
\end{figure}

This reasoning by itself suggests simply that configurations appropriate to the temperature of interest should be used to scale to higher densities. However, recall that the loss curves fail to collapse at the lowest values of $T_\text{ref}$, and here the configurations are indeed appropriate (for the lowest temperature the TM-method is the same as the Q-method). For additional insight we consider that a necessary condition for the force method to be applicable is that the force vectors before and after rescaling are parallel. This condition will always be fulfilled only approximately. In Fig.~\ref{fig:forceangles}, we plot the distributions of the cosine of the angle $\alpha$ between the forces before and after rescaling for all 8000 particles in the system for different densities, both for configurations drawn from $T_\text{ref}=0.2$ and from $T=0.6$ (i.e., appropriate for the TM-method at $T_\text{ref}=0.6$), at density $\rho=1.3$. In the ideal case, we would obtain a vertical line at value one. We see that with larger densities, the angle between forces also tends to increase. Thus independent of whether they are to be used at a (reference) temperature of 0.2 or at a higher one, these configurations scale less perfectly once the density is increased to 2.0. They have generally poorer isomorph-scaling qualities. This presumably explains the failure of the loss curves to collapse at low $T_\text{ref}$.

%

To quantify this statement, we introduce a measure for the misalignment between the forces before and after rescaling, given by
\begin{equation}
\frac{1}{N}\sum\limits_{i=1}^{N} (1-\cos(\alpha_i))/N\;,
\label{eq:measure}
\end{equation}
with $i$ counting through all particles up to the number of particles $N$ is the system. In table~\ref{tab:misalignment}, we present the obtained values. These values also suggest that the TM-method is more suitable to determine isomorphic temperatures via the force method, because the relevant configurations have better density scaling properties..

\begin{figure}
\centering
      \includegraphics[width=0.5\textwidth]{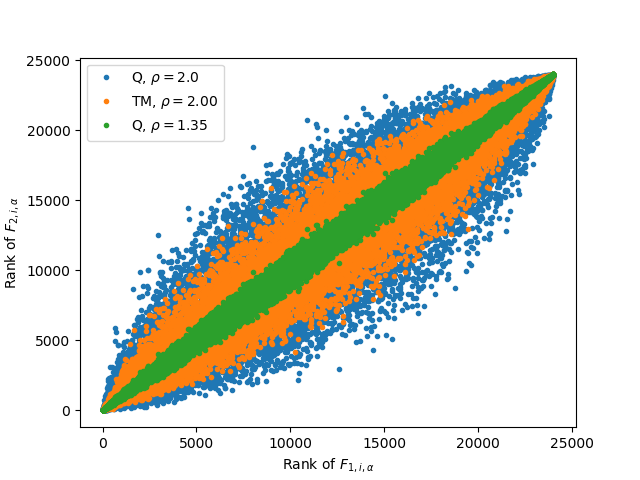}
      \caption{Ranks of force vector components before ($F_1$) and after ($F_2$) rescaling from a configuration with density $\rho=1.30$ to the same configurations considered in Fig.~\ref{fig:forceangles}. The indices $i,\alpha$ refer to the number of the particle and the spatial coordinate, respectively. The hierarchy of correlation strengths illustrated here matches the values stated in Fig.~\ref{fig:Rplot}.}
      \label{fig:Spearman}
\end{figure}

Additionally, we consider the Spearman correlations~\cite{Spearman:1904,Kendall/Gibbons:1990} of the force vectors before and after rescaling as in~\cite{Schroder2021}, where it was suggested that good isomophs have Spearman correlations larger than $~0.95$. In Fig.~\ref{fig:Rplot} the Spearman correlations for the configurations drawn from the cooling run at that indicated temperatures and scaled from the initial density 1.30 to the other densities considered in this study. In agreement with the non-collapse of loss curves for $\rho=2.00$ with the Q-method in Fig.~\ref{fig:reduced}, we find that this configuration does not satisfy the suggested criterion--the lowest temperature point of the $\rho=2.00$ curve lies below 0.95. Such low-temperature configurations were used for all temperatures in the Q-method. As an additional illustration, we present in Fig.~\ref{fig:Spearman} the ranks of the force vectors before and after rescaling for the same configurations as in Fig.~\ref{fig:forceangles}. The Spearman correlation analysis emphasizes that the configurations obtained at the lowest temperatures at the initial reference density are intrinsically poorly scaling when large density changes are considered.

\begin{table}
  \centering
  \sisetup {
    per-mode = fraction
  }
  \begin{tabular}{c|ccc}
\toprule
Misalign.&$\rho=1.35$&$\rho=2.00$\\
\hline
Quenched&\SI{3.2e-3}{}&\SI{8.8e-2}{}\\
Temperature-matched&\SI{1.0e-3}{}&\SI{3.6e-2}{}\\
\hline
  \end{tabular}
  \caption{Measure~\eqref{eq:measure} for the misalignment of forces before and after rescaling. The lower the value, the more aligned the forces and the more suitable the force method in order to determine isomorphic temperatures.}
\label{tab:misalignment}
\end{table}
\section{Conclusion}
\label{sec:discussion_conclusions}

Our results show that the isomorph theory makes successful predictions for the behavior of an R-simple glassy system in the context of dynamic mechanical analysis. For a change in density a corresponding temperature change and thereby a corresponding applied frequency can be identified such that the loss curve is invariant when compared in reduced units. The factor by which temperature should be changed can be identified by scaling configurations and comparing the force vectors of the scaled and unscaled configurations. For moderate density changes it is sufficient to consider the same configuration independent of which temperature the DMA was carried out; this leads to temperature ratios independent of reference temperatures, while for large density changes, up to 50\% increase, it was found important to apply the scaling procedure to configurations more typical of the corresponding temperature. In this case the temperature ratios for different density changes depend slightly on the starting temperature ($T_\text{ref}$).

Even with the correct choice of configurations to use for determining the isomorphic temperatures there is an incomplete collapse at the highest density at the low-temperature end of the response. Physically, this deviation corresponds to excess loss at high density, compared to when the same glass is probed in essentially the same manner at lower densities (and correspondingly lower temperatures). A possible interpretation, which is purely speculative, is that the low-temperature excess loss could be related to a Johari-Goldstein beta process. A detailed investigation of the atomic motions involved and their differences at the different densities, beyond the scope of the present work, will be required to clarify this.


Like density scaling in the case of glass-forming liquids~\cite{Tolle/others:1998,Alba-Simionesco/others:2004,Casalini/Roland:2004}, our results highlight the importance of density rather than pressure as a more fundamental thermodynamic parameter for the understanding of R-simple systems' behavior. It would be interesting to see the prediction of isomorphic DMA profiles tested experimentally. This would be very challenging for metallic glasses due to the high pressures required, but could perhaps be carried out for small organic molecules or polymers. There is no experimental analog for the force method, so a more empirical protocol must be used. For example, assuming moderate density changes one could make the assumption that the temperature factors associated with a given density are independent of the starting temperature $T_\text{ref}$. If we ignore for the moment the experimental difficulty in controlling density one proceeds as follows: Given two densities $\rho_i$ and $\rho_f$ to be compared, (1) choose a particular temperature $T_{0,i}$ at the starting density $\rho_i$; (2) identify empirically the temperature $T_{0,f}$ at which the DMA loss in reduced units $\tilde G''$ at the state point $\rho_f, T_{0,f}$ is equal to the reduced loss at state point $\rho_i, T_{0,i}$ where it is understood that the frequency is adjusted to maintain a fixed reduced value; (3) for other temperatures at the initial density $\rho_i$ apply the same temperature factor $T_{f,0}/T_{i,0}$ to determine the isomorphic state points at density $\rho_f$. In this way one empirically found parameter, $T_{f,0}/T_{i,0}$ is sufficient to predict the complete temperature-dependent loss curve for the new density $\rho_f$. The procedure must then be repeated for other densities. With sufficiently accurate equation of state data it should be possible to adapt this protocol to the case of DMA curves determined at fixed pressure rather than fixed density, although we have not considered in detail how this might be done. In any case one should be aware that a fixed frequency protocol at a given pressure will in general not map to a fixed frequency at a different pressure.

\section{Data availability}

The data needed to reproduce most of the figures in this paper are available at http://doi.org/10.5281/zenodo.6542884.
\begin{acknowledgments}
We thank Jeppe Dyre for reading the manuscript and giving useful suggestions. The work was supported in part by the VILLUM Foundation's {\em
Matter} Grant (No. 16515) and the Deutsche Forschungsgemeinschaft
Grant (No. 461147152).
\end{acknowledgments}

\bibliography{literature}

\begin{thebibliography}{45}%
\makeatletter
\providecommand \@ifxundefined [1]{%
 \@ifx{#1\undefined}
}%
\providecommand \@ifnum [1]{%
 \ifnum #1\expandafter \@firstoftwo
 \else \expandafter \@secondoftwo
 \fi
}%
\providecommand \@ifx [1]{%
 \ifx #1\expandafter \@firstoftwo
 \else \expandafter \@secondoftwo
 \fi
}%
\providecommand \natexlab [1]{#1}%
\providecommand \enquote  [1]{``#1''}%
\providecommand \bibnamefont  [1]{#1}%
\providecommand \bibfnamefont [1]{#1}%
\providecommand \citenamefont [1]{#1}%
\providecommand \href@noop [0]{\@secondoftwo}%
\providecommand \href [0]{\begingroup \@sanitize@url \@href}%
\providecommand \@href[1]{\@@startlink{#1}\@@href}%
\providecommand \@@href[1]{\endgroup#1\@@endlink}%
\providecommand \@sanitize@url [0]{\catcode `\\12\catcode `\$12\catcode
  `\&12\catcode `\#12\catcode `\^12\catcode `\_12\catcode `\%12\relax}%
\providecommand \@@startlink[1]{}%
\providecommand \@@endlink[0]{}%
\providecommand \url  [0]{\begingroup\@sanitize@url \@url }%
\providecommand \@url [1]{\endgroup\@href {#1}{\urlprefix }}%
\providecommand \urlprefix  [0]{URL }%
\providecommand \Eprint [0]{\href }%
\providecommand \doibase [0]{https://doi.org/}%
\providecommand \selectlanguage [0]{\@gobble}%
\providecommand \bibinfo  [0]{\@secondoftwo}%
\providecommand \bibfield  [0]{\@secondoftwo}%
\providecommand \translation [1]{[#1]}%
\providecommand \BibitemOpen [0]{}%
\providecommand \bibitemStop [0]{}%
\providecommand \bibitemNoStop [0]{.\EOS\space}%
\providecommand \EOS [0]{\spacefactor3000\relax}%
\providecommand \BibitemShut  [1]{\csname bibitem#1\endcsname}%
\let\auto@bib@innerbib\@empty
\bibitem [{\citenamefont {Eckert}\ \emph {et~al.}(2007)\citenamefont {Eckert},
  \citenamefont {Das}, \citenamefont {Pauly},\ and\ \citenamefont
  {Duhamel}}]{Eckert2007}%
  \BibitemOpen
  \bibfield  {author} {\bibinfo {author} {\bibfnamefont {J.}~\bibnamefont
  {Eckert}}, \bibinfo {author} {\bibfnamefont {J.}~\bibnamefont {Das}},
  \bibinfo {author} {\bibfnamefont {S.}~\bibnamefont {Pauly}},\ and\ \bibinfo
  {author} {\bibfnamefont {C.}~\bibnamefont {Duhamel}},\ }\bibfield  {title}
  {\bibinfo {title} {Mechanical properties of bulk metallic glasses and
  composites},\ }\href {https://doi.org/10.1557/jmr.2007.0050} {\bibfield
  {journal} {\bibinfo  {journal} {J. of Mater. Res.}\ }\textbf {\bibinfo
  {volume} {22}},\ \bibinfo {pages} {285} (\bibinfo {year} {2007})}\BibitemShut
  {NoStop}%
\bibitem [{\citenamefont {Liu}\ \emph {et~al.}(2015)\citenamefont {Liu},
  \citenamefont {Pineda},\ and\ \citenamefont {Crespo}}]{Liu2015}%
  \BibitemOpen
  \bibfield  {author} {\bibinfo {author} {\bibfnamefont {C.}~\bibnamefont
  {Liu}}, \bibinfo {author} {\bibfnamefont {E.}~\bibnamefont {Pineda}},\ and\
  \bibinfo {author} {\bibfnamefont {D.}~\bibnamefont {Crespo}},\ }\bibfield
  {title} {\bibinfo {title} {Mechanical relaxation of metallic glasses: An
  overview of experimental data and theoretical models},\ }\href
  {https://doi.org/10.3390/met5021073} {\bibfield  {journal} {\bibinfo
  {journal} {Metals}\ }\textbf {\bibinfo {volume} {5}},\ \bibinfo {pages}
  {1073} (\bibinfo {year} {2015})}\BibitemShut {NoStop}%
\bibitem [{\citenamefont {Hand}\ and\ \citenamefont
  {Tadjiev}(2010)}]{Hand2010}%
  \BibitemOpen
  \bibfield  {author} {\bibinfo {author} {\bibfnamefont {R.~J.}\ \bibnamefont
  {Hand}}\ and\ \bibinfo {author} {\bibfnamefont {D.~R.}\ \bibnamefont
  {Tadjiev}},\ }\bibfield  {title} {\bibinfo {title} {Mechanical properties of
  silicate glasses as a function of composition},\ }\href
  {https://doi.org/10.1016/j.jnoncrysol.2010.05.007} {\bibfield  {journal}
  {\bibinfo  {journal} {J. of Non-Cryst. Solids}\ }\textbf {\bibinfo {volume}
  {356}},\ \bibinfo {pages} {2417} (\bibinfo {year} {2010})}\BibitemShut
  {NoStop}%
\bibitem [{\citenamefont {Louzguine-Luzgin}\ \emph {et~al.}(2012)\citenamefont
  {Louzguine-Luzgin}, \citenamefont {Louzguina-Luzgina},\ and\ \citenamefont
  {Churyumov}}]{LouzguineLuzgin2012}%
  \BibitemOpen
  \bibfield  {author} {\bibinfo {author} {\bibfnamefont {D.}~\bibnamefont
  {Louzguine-Luzgin}}, \bibinfo {author} {\bibfnamefont {L.}~\bibnamefont
  {Louzguina-Luzgina}},\ and\ \bibinfo {author} {\bibfnamefont
  {A.}~\bibnamefont {Churyumov}},\ }\bibfield  {title} {\bibinfo {title}
  {Mechanical properties and deformation behavior of bulk metallic glasses},\
  }\href {https://doi.org/10.3390/met3010001} {\bibfield  {journal} {\bibinfo
  {journal} {Metals}\ }\textbf {\bibinfo {volume} {3}},\ \bibinfo {pages} {1}
  (\bibinfo {year} {2012})}\BibitemShut {NoStop}%
\bibitem [{\citenamefont {Moln{\'{a}}r}\ \emph {et~al.}(2017)\citenamefont
  {Moln{\'{a}}r}, \citenamefont {Ganster},\ and\ \citenamefont
  {Tanguy}}]{Molnr2017}%
  \BibitemOpen
  \bibfield  {author} {\bibinfo {author} {\bibfnamefont {G.}~\bibnamefont
  {Moln{\'{a}}r}}, \bibinfo {author} {\bibfnamefont {P.}~\bibnamefont
  {Ganster}},\ and\ \bibinfo {author} {\bibfnamefont {A.}~\bibnamefont
  {Tanguy}},\ }\bibfield  {title} {\bibinfo {title} {Effect of composition and
  pressure on the shear strength of sodium silicate glasses: An atomic scale
  simulation study},\ }\bibfield  {journal} {\bibinfo  {journal} {Phys. Rev.
  E}\ }\textbf {\bibinfo {volume} {95}},\ \href
  {https://doi.org/10.1103/physreve.95.043001} {10.1103/physreve.95.043001}
  (\bibinfo {year} {2017})\BibitemShut {NoStop}%
\bibitem [{\citenamefont {T\"olle}\ \emph {et~al.}(1998)\citenamefont
  {T\"olle}, \citenamefont {Schober}, \citenamefont {Wuttke}, \citenamefont
  {Randl},\ and\ \citenamefont {Fujara}}]{Tolle/others:1998}%
  \BibitemOpen
  \bibfield  {author} {\bibinfo {author} {\bibfnamefont {A.}~\bibnamefont
  {T\"olle}}, \bibinfo {author} {\bibfnamefont {H.}~\bibnamefont {Schober}},
  \bibinfo {author} {\bibfnamefont {J.}~\bibnamefont {Wuttke}}, \bibinfo
  {author} {\bibfnamefont {O.}~\bibnamefont {Randl}},\ and\ \bibinfo {author}
  {\bibfnamefont {F.}~\bibnamefont {Fujara}},\ }\bibfield  {title} {\bibinfo
  {title} {{Fast relaxation in a fragile liquid under pressure}},\ }\href
  {https://doi.org/{10.1103/PhysRevLett.80.2374}} {\bibfield  {journal}
  {\bibinfo  {journal} {Phys. Rev. Lett.}\ }\textbf {\bibinfo {volume}
  {{80}}},\ \bibinfo {pages} {2374} (\bibinfo {year} {1998})}\BibitemShut
  {NoStop}%
\bibitem [{\citenamefont {Alba-Simionesco}\ \emph {et~al.}(2004)\citenamefont
  {Alba-Simionesco}, \citenamefont {Cailliaux}, \citenamefont {Alegria},\ and\
  \citenamefont {Tarjus}}]{Alba-Simionesco/others:2004}%
  \BibitemOpen
  \bibfield  {author} {\bibinfo {author} {\bibfnamefont {C.}~\bibnamefont
  {Alba-Simionesco}}, \bibinfo {author} {\bibfnamefont {A.}~\bibnamefont
  {Cailliaux}}, \bibinfo {author} {\bibfnamefont {A.}~\bibnamefont {Alegria}},\
  and\ \bibinfo {author} {\bibfnamefont {G.}~\bibnamefont {Tarjus}},\
  }\bibfield  {title} {\bibinfo {title} {{Scaling out the density dependence of
  the $\alpha$-relaxation in glass-forming polymers}},\ }\href
  {https://doi.org/10.1209/epl/i2004-10214-6} {\bibfield  {journal} {\bibinfo
  {journal} {Europhys. Lett.}\ }\textbf {\bibinfo {volume} {68}},\ \bibinfo
  {pages} {58} (\bibinfo {year} {2004})}\BibitemShut {NoStop}%
\bibitem [{\citenamefont {Casalini}\ and\ \citenamefont
  {Roland}(2004)}]{Casalini/Roland:2004}%
  \BibitemOpen
  \bibfield  {author} {\bibinfo {author} {\bibfnamefont {R.}~\bibnamefont
  {Casalini}}\ and\ \bibinfo {author} {\bibfnamefont {C.~M.}\ \bibnamefont
  {Roland}},\ }\bibfield  {title} {\bibinfo {title} {Thermodynamical scaling of
  the glass transition dynamics},\ }\href
  {https://doi.org/10.1103/PhysRevE.69.062501} {\bibfield  {journal} {\bibinfo
  {journal} {Phys. Rev. E}\ }\textbf {\bibinfo {volume} {69}},\ \bibinfo
  {pages} {062501} (\bibinfo {year} {2004})}\BibitemShut {NoStop}%
\bibitem [{\citenamefont {B{\o}hling}\ \emph {et~al.}(2012)\citenamefont
  {B{\o}hling}, \citenamefont {Ingebrigtsen}, \citenamefont {Grzybowski},
  \citenamefont {Paluch}, \citenamefont {Dyre},\ and\ \citenamefont
  {Schr{\o}der}}]{Boehling2012}%
  \BibitemOpen
  \bibfield  {author} {\bibinfo {author} {\bibfnamefont {L.}~\bibnamefont
  {B{\o}hling}}, \bibinfo {author} {\bibfnamefont {T.~S.}\ \bibnamefont
  {Ingebrigtsen}}, \bibinfo {author} {\bibfnamefont {A.}~\bibnamefont
  {Grzybowski}}, \bibinfo {author} {\bibfnamefont {M.}~\bibnamefont {Paluch}},
  \bibinfo {author} {\bibfnamefont {J.~C.}\ \bibnamefont {Dyre}},\ and\
  \bibinfo {author} {\bibfnamefont {T.~B.}\ \bibnamefont {Schr{\o}der}},\
  }\bibfield  {title} {\bibinfo {title} {Scaling of viscous dynamics in simple
  liquids: theory, simulation and experiment},\ }\href
  {https://doi.org/10.1088/1367-2630/14/11/113035} {\bibfield  {journal}
  {\bibinfo  {journal} {New J. Phys.}\ }\textbf {\bibinfo {volume} {14}},\
  \bibinfo {pages} {113035} (\bibinfo {year} {2012})}\BibitemShut {NoStop}%
\bibitem [{\citenamefont {Sanz}\ \emph {et~al.}(2019)\citenamefont {Sanz},
  \citenamefont {Hecksher}, \citenamefont {Hansen}, \citenamefont {Dyre},
  \citenamefont {Niss},\ and\ \citenamefont {Pedersen}}]{Sanz/Others:2019}%
  \BibitemOpen
  \bibfield  {author} {\bibinfo {author} {\bibfnamefont {A.}~\bibnamefont
  {Sanz}}, \bibinfo {author} {\bibfnamefont {T.}~\bibnamefont {Hecksher}},
  \bibinfo {author} {\bibfnamefont {H.~W.}\ \bibnamefont {Hansen}}, \bibinfo
  {author} {\bibfnamefont {J.~C.}\ \bibnamefont {Dyre}}, \bibinfo {author}
  {\bibfnamefont {K.}~\bibnamefont {Niss}},\ and\ \bibinfo {author}
  {\bibfnamefont {U.~R.}\ \bibnamefont {Pedersen}},\ }\bibfield  {title}
  {\bibinfo {title} {Experimental evidence for a state-point-dependent
  density-scaling exponent of liquid dynamics},\ }\href
  {https://doi.org/10.1103/PhysRevLett.122.055501} {\bibfield  {journal}
  {\bibinfo  {journal} {Phys. Rev. Lett.}\ }\textbf {\bibinfo {volume} {122}},\
  \bibinfo {pages} {055501} (\bibinfo {year} {2019})}\BibitemShut {NoStop}%
\bibitem [{\citenamefont {Casalini}\ and\ \citenamefont
  {Ransom}(2020)}]{Casalini/Ransom:2020}%
  \BibitemOpen
  \bibfield  {author} {\bibinfo {author} {\bibfnamefont {R.}~\bibnamefont
  {Casalini}}\ and\ \bibinfo {author} {\bibfnamefont {T.~C.}\ \bibnamefont
  {Ransom}},\ }\bibfield  {title} {\bibinfo {title} {On the pressure dependence
  of the thermodynamical scaling exponent $\gamma$},\ }\href
  {https://doi.org/10.1039/d0sm00254b} {\bibfield  {journal} {\bibinfo
  {journal} {Soft Matter}\ }\textbf {\bibinfo {volume} {16}},\ \bibinfo {pages}
  {4625} (\bibinfo {year} {2020})}\BibitemShut {NoStop}%
\bibitem [{\citenamefont {Bailey}\ \emph
  {et~al.}(2008{\natexlab{a}})\citenamefont {Bailey}, \citenamefont {Pedersen},
  \citenamefont {Gnan}, \citenamefont {Schr{\o}der},\ and\ \citenamefont
  {Dyre}}]{Bailey2008a}%
  \BibitemOpen
  \bibfield  {author} {\bibinfo {author} {\bibfnamefont {N.~P.}\ \bibnamefont
  {Bailey}}, \bibinfo {author} {\bibfnamefont {U.~R.}\ \bibnamefont
  {Pedersen}}, \bibinfo {author} {\bibfnamefont {N.}~\bibnamefont {Gnan}},
  \bibinfo {author} {\bibfnamefont {T.~B.}\ \bibnamefont {Schr{\o}der}},\ and\
  \bibinfo {author} {\bibfnamefont {J.~C.}\ \bibnamefont {Dyre}},\ }\bibfield
  {title} {\bibinfo {title} {{Pressure-energy correlations in liquids. I.
  Results from computer simulations}},\ }\href
  {https://doi.org/10.1063/1.2982247} {\bibfield  {journal} {\bibinfo
  {journal} {J. Chem. Phys.}\ }\textbf {\bibinfo {volume} {129}},\ \bibinfo
  {pages} {184507} (\bibinfo {year} {2008}{\natexlab{a}})}\BibitemShut
  {NoStop}%
\bibitem [{\citenamefont {Bailey}\ \emph
  {et~al.}(2008{\natexlab{b}})\citenamefont {Bailey}, \citenamefont {Pedersen},
  \citenamefont {Gnan}, \citenamefont {Schr{\o}der},\ and\ \citenamefont
  {Dyre}}]{Bailey2008b}%
  \BibitemOpen
  \bibfield  {author} {\bibinfo {author} {\bibfnamefont {N.~P.}\ \bibnamefont
  {Bailey}}, \bibinfo {author} {\bibfnamefont {U.~R.}\ \bibnamefont
  {Pedersen}}, \bibinfo {author} {\bibfnamefont {N.}~\bibnamefont {Gnan}},
  \bibinfo {author} {\bibfnamefont {T.~B.}\ \bibnamefont {Schr{\o}der}},\ and\
  \bibinfo {author} {\bibfnamefont {J.~C.}\ \bibnamefont {Dyre}},\ }\bibfield
  {title} {\bibinfo {title} {Pressure-energy correlations in liquids. {II}.
  analysis and consequences},\ }\href {https://doi.org/10.1063/1.2982249}
  {\bibfield  {journal} {\bibinfo  {journal} {J. Chem. Phys.}\ }\textbf
  {\bibinfo {volume} {129}},\ \bibinfo {pages} {184508} (\bibinfo {year}
  {2008}{\natexlab{b}})}\BibitemShut {NoStop}%
\bibitem [{\citenamefont {Schr{\o}der}\ \emph {et~al.}(2009)\citenamefont
  {Schr{\o}der}, \citenamefont {Bailey}, \citenamefont {Pedersen},
  \citenamefont {Gnan},\ and\ \citenamefont {Dyre}}]{Schroder2009}%
  \BibitemOpen
  \bibfield  {author} {\bibinfo {author} {\bibfnamefont {T.~B.}\ \bibnamefont
  {Schr{\o}der}}, \bibinfo {author} {\bibfnamefont {N.~P.}\ \bibnamefont
  {Bailey}}, \bibinfo {author} {\bibfnamefont {U.~R.}\ \bibnamefont
  {Pedersen}}, \bibinfo {author} {\bibfnamefont {N.}~\bibnamefont {Gnan}},\
  and\ \bibinfo {author} {\bibfnamefont {J.~C.}\ \bibnamefont {Dyre}},\
  }\bibfield  {title} {\bibinfo {title} {Pressure-energy correlations in
  liquids. {III}. statistical mechanics and thermodynamics of liquids with
  hidden scale invariance},\ }\href {https://doi.org/10.1063/1.3265955}
  {\bibfield  {journal} {\bibinfo  {journal} {J. Chem. Phys.}\ }\textbf
  {\bibinfo {volume} {131}},\ \bibinfo {pages} {234503} (\bibinfo {year}
  {2009})}\BibitemShut {NoStop}%
\bibitem [{\citenamefont {Gnan}\ \emph {et~al.}(2009)\citenamefont {Gnan},
  \citenamefont {Schr{\o}der}, \citenamefont {Pedersen}, \citenamefont
  {Bailey},\ and\ \citenamefont {Dyre}}]{Gnan2009}%
  \BibitemOpen
  \bibfield  {author} {\bibinfo {author} {\bibfnamefont {N.}~\bibnamefont
  {Gnan}}, \bibinfo {author} {\bibfnamefont {T.~B.}\ \bibnamefont
  {Schr{\o}der}}, \bibinfo {author} {\bibfnamefont {U.~R.}\ \bibnamefont
  {Pedersen}}, \bibinfo {author} {\bibfnamefont {N.~P.}\ \bibnamefont
  {Bailey}},\ and\ \bibinfo {author} {\bibfnamefont {J.~C.}\ \bibnamefont
  {Dyre}},\ }\bibfield  {title} {\bibinfo {title} {{Pressure-energy
  correlations in liquids. IV. {\textquotedblleft}Isomorphs{\textquotedblright}
  in liquid phase diagrams}},\ }\href {https://doi.org/10.1063/1.3265957}
  {\bibfield  {journal} {\bibinfo  {journal} {J. Chem. Phys.}\ }\textbf
  {\bibinfo {volume} {131}},\ \bibinfo {pages} {234504} (\bibinfo {year}
  {2009})}\BibitemShut {NoStop}%
\bibitem [{\citenamefont {Schr{\o}der}\ \emph {et~al.}(2011)\citenamefont
  {Schr{\o}der}, \citenamefont {Gnan}, \citenamefont {Pedersen}, \citenamefont
  {Bailey},\ and\ \citenamefont {Dyre}}]{Schroder2011}%
  \BibitemOpen
  \bibfield  {author} {\bibinfo {author} {\bibfnamefont {T.~B.}\ \bibnamefont
  {Schr{\o}der}}, \bibinfo {author} {\bibfnamefont {N.}~\bibnamefont {Gnan}},
  \bibinfo {author} {\bibfnamefont {U.~R.}\ \bibnamefont {Pedersen}}, \bibinfo
  {author} {\bibfnamefont {N.~P.}\ \bibnamefont {Bailey}},\ and\ \bibinfo
  {author} {\bibfnamefont {J.~C.}\ \bibnamefont {Dyre}},\ }\bibfield  {title}
  {\bibinfo {title} {Pressure-energy correlations in liquids. v. isomorphs in
  generalized {Lennard-Jones} systems},\ }\href
  {https://doi.org/10.1063/1.3582900} {\bibfield  {journal} {\bibinfo
  {journal} {J. Chem. Phys.}\ }\textbf {\bibinfo {volume} {134}},\ \bibinfo
  {pages} {164505} (\bibinfo {year} {2011})}\BibitemShut {NoStop}%
\bibitem [{\citenamefont {Ingebrigtsen}\ \emph {et~al.}(2012)\citenamefont
  {Ingebrigtsen}, \citenamefont {B{\o}hling}, \citenamefont {Schr{\o}der},\
  and\ \citenamefont {Dyre}}]{Ingebrigtsen2012}%
  \BibitemOpen
  \bibfield  {author} {\bibinfo {author} {\bibfnamefont {T.~S.}\ \bibnamefont
  {Ingebrigtsen}}, \bibinfo {author} {\bibfnamefont {L.}~\bibnamefont
  {B{\o}hling}}, \bibinfo {author} {\bibfnamefont {T.~B.}\ \bibnamefont
  {Schr{\o}der}},\ and\ \bibinfo {author} {\bibfnamefont {J.~C.}\ \bibnamefont
  {Dyre}},\ }\bibfield  {title} {\bibinfo {title} {Communication:
  {Thermodynamics} of condensed matter with strong pressure-energy
  correlations},\ }\href {https://doi.org/10.1063/1.3685804} {\bibfield
  {journal} {\bibinfo  {journal} {J. Chem. Phys.}\ }\textbf {\bibinfo {volume}
  {136}},\ \bibinfo {pages} {061102} (\bibinfo {year} {2012})}\BibitemShut
  {NoStop}%
\bibitem [{\citenamefont {Schr{\o}der}\ and\ \citenamefont
  {Dyre}(2014)}]{Schroder2014}%
  \BibitemOpen
  \bibfield  {author} {\bibinfo {author} {\bibfnamefont {T.~B.}\ \bibnamefont
  {Schr{\o}der}}\ and\ \bibinfo {author} {\bibfnamefont {J.~C.}\ \bibnamefont
  {Dyre}},\ }\bibfield  {title} {\bibinfo {title} {Simplicity of condensed
  matter at its core: Generic definition of a {Roskilde-simple} system},\
  }\href {https://doi.org/10.1063/1.4901215} {\bibfield  {journal} {\bibinfo
  {journal} {J. Chem. Phys.}\ }\textbf {\bibinfo {volume} {141}},\ \bibinfo
  {pages} {204502} (\bibinfo {year} {2014})}\BibitemShut {NoStop}%
\bibitem [{\citenamefont {Dyre}(2014)}]{Dyre2014}%
  \BibitemOpen
  \bibfield  {author} {\bibinfo {author} {\bibfnamefont {J.~C.}\ \bibnamefont
  {Dyre}},\ }\bibfield  {title} {\bibinfo {title} {Hidden scale invariance in
  condensed matter},\ }\href {https://doi.org/10.1021/jp501852b} {\bibfield
  {journal} {\bibinfo  {journal} {J. Phys. Chem. B}\ }\textbf {\bibinfo
  {volume} {118}},\ \bibinfo {pages} {10007} (\bibinfo {year}
  {2014})}\BibitemShut {NoStop}%
\bibitem [{\citenamefont {R{\"o}sner}\ \emph {et~al.}(2004)\citenamefont
  {R{\"o}sner}, \citenamefont {Samwer},\ and\ \citenamefont
  {Lunkenheimer}}]{Rosner/Samwer/Lunkenheimer:2004}%
  \BibitemOpen
  \bibfield  {author} {\bibinfo {author} {\bibfnamefont {P.}~\bibnamefont
  {R{\"o}sner}}, \bibinfo {author} {\bibfnamefont {K.}~\bibnamefont {Samwer}},\
  and\ \bibinfo {author} {\bibfnamefont {P.}~\bibnamefont {Lunkenheimer}},\
  }\bibfield  {title} {\bibinfo {title} {Indications for an ``excess wing'' in
  metallic glasses from the mechanical loss modulus in
  {Zr}$_{65}${Al}$_{7.5}${Cu}$_{27.5}$},\ }\href
  {https://doi.org/10.1209/epl/i2004-10193-6} {\bibfield  {journal} {\bibinfo
  {journal} {EPL}\ }\textbf {\bibinfo {volume} {68}},\ \bibinfo {pages} {226}
  (\bibinfo {year} {2004})}\BibitemShut {NoStop}%
\bibitem [{\citenamefont {Galloway}\ \emph {et~al.}(2020)\citenamefont
  {Galloway}, \citenamefont {Ma}, \citenamefont {Keim}, \citenamefont
  {Jerolmack}, \citenamefont {Yodh},\ and\ \citenamefont
  {Arratia}}]{Galloway2020}%
  \BibitemOpen
  \bibfield  {author} {\bibinfo {author} {\bibfnamefont {K.~L.}\ \bibnamefont
  {Galloway}}, \bibinfo {author} {\bibfnamefont {X.}~\bibnamefont {Ma}},
  \bibinfo {author} {\bibfnamefont {N.~C.}\ \bibnamefont {Keim}}, \bibinfo
  {author} {\bibfnamefont {D.~J.}\ \bibnamefont {Jerolmack}}, \bibinfo {author}
  {\bibfnamefont {A.~G.}\ \bibnamefont {Yodh}},\ and\ \bibinfo {author}
  {\bibfnamefont {P.~E.}\ \bibnamefont {Arratia}},\ }\bibfield  {title}
  {\bibinfo {title} {Scaling of relaxation and excess entropy in plastically
  deformed amorphous solids},\ }\href {https://doi.org/10.1073/pnas.2000698117}
  {\bibfield  {journal} {\bibinfo  {journal} {PNAS}\ }\textbf {\bibinfo
  {volume} {117}},\ \bibinfo {pages} {11887} (\bibinfo {year}
  {2020})}\BibitemShut {NoStop}%
\bibitem [{\citenamefont {Cohen}\ \emph {et~al.}(2012)\citenamefont {Cohen},
  \citenamefont {Karmakar}, \citenamefont {Procaccia},\ and\ \citenamefont
  {Samwer}}]{Cohen2012}%
  \BibitemOpen
  \bibfield  {author} {\bibinfo {author} {\bibfnamefont {Y.}~\bibnamefont
  {Cohen}}, \bibinfo {author} {\bibfnamefont {S.}~\bibnamefont {Karmakar}},
  \bibinfo {author} {\bibfnamefont {I.}~\bibnamefont {Procaccia}},\ and\
  \bibinfo {author} {\bibfnamefont {K.}~\bibnamefont {Samwer}},\ }\bibfield
  {title} {\bibinfo {title} {The nature of the $\beta$-peak in the loss modulus
  of amorphous solids},\ }\href {https://doi.org/10.1209/0295-5075/100/36003}
  {\bibfield  {journal} {\bibinfo  {journal} {EPL}\ }\textbf {\bibinfo {volume}
  {100}},\ \bibinfo {pages} {36003} (\bibinfo {year} {2012})}\BibitemShut
  {NoStop}%
\bibitem [{\citenamefont {Yu}\ and\ \citenamefont
  {Samwer}(2014)}]{Yu/Samwer:2014}%
  \BibitemOpen
  \bibfield  {author} {\bibinfo {author} {\bibfnamefont {H.-B.}\ \bibnamefont
  {Yu}}\ and\ \bibinfo {author} {\bibfnamefont {K.}~\bibnamefont {Samwer}},\
  }\bibfield  {title} {\bibinfo {title} {{Atomic mechanism of internal friction
  in a model metallic glass}},\ }\href
  {https://doi.org/{10.1103/PhysRevB.90.144201}} {\bibfield  {journal}
  {\bibinfo  {journal} {Phys. Rev. B}\ }\textbf {\bibinfo {volume} {90}},\
  \bibinfo {pages} {144201} (\bibinfo {year} {2014})}\BibitemShut {NoStop}%
\bibitem [{\citenamefont {Yu}\ \emph {et~al.}(2015)\citenamefont {Yu},
  \citenamefont {Richert}, \citenamefont {Maass},\ and\ \citenamefont
  {Samwer}}]{Yu/others:2015}%
  \BibitemOpen
  \bibfield  {author} {\bibinfo {author} {\bibfnamefont {H.-B.}\ \bibnamefont
  {Yu}}, \bibinfo {author} {\bibfnamefont {R.}~\bibnamefont {Richert}},
  \bibinfo {author} {\bibfnamefont {R.}~\bibnamefont {Maass}},\ and\ \bibinfo
  {author} {\bibfnamefont {K.}~\bibnamefont {Samwer}},\ }\bibfield  {title}
  {\bibinfo {title} {{Strain induced fragility transition in metallic glass}},\
  }\href {https://doi.org/{10.1038/ncomms8179}} {\bibfield  {journal} {\bibinfo
   {journal} {Nat. Commun.}\ }\textbf {\bibinfo {volume} {6}},\ \bibinfo
  {pages} {7179} (\bibinfo {year} {2015})}\BibitemShut {NoStop}%
\bibitem [{\citenamefont {Yu}\ \emph {et~al.}(2017)\citenamefont {Yu},
  \citenamefont {Richert},\ and\ \citenamefont
  {Samwer}}]{Yu/Richert/Samwer:2017}%
  \BibitemOpen
  \bibfield  {author} {\bibinfo {author} {\bibfnamefont {H.}~\bibnamefont
  {Yu}}, \bibinfo {author} {\bibfnamefont {R.}~\bibnamefont {Richert}},\ and\
  \bibinfo {author} {\bibfnamefont {K.}~\bibnamefont {Samwer}},\ }\bibfield
  {title} {\bibinfo {title} {{Structural rearrangements governing
  Johari-Goldstein relaxations in metallic glasses}},\ }\href
  {https://doi.org/10.1126/sciadv.1701577} {\bibfield  {journal} {\bibinfo
  {journal} {Sci. Adv.}\ }\textbf {\bibinfo {volume} {3}},\ \bibinfo {pages}
  {e1701577} (\bibinfo {year} {2017})}\BibitemShut {NoStop}%
\bibitem [{\citenamefont {Bailey}\ \emph {et~al.}(2013)\citenamefont {Bailey},
  \citenamefont {B{\o}hling}, \citenamefont {Veldhorst}, \citenamefont
  {Schr{\o}der},\ and\ \citenamefont {Dyre}}]{Bailey/Others:2013}%
  \BibitemOpen
  \bibfield  {author} {\bibinfo {author} {\bibfnamefont {N.~P.}\ \bibnamefont
  {Bailey}}, \bibinfo {author} {\bibfnamefont {L.}~\bibnamefont {B{\o}hling}},
  \bibinfo {author} {\bibfnamefont {A.~A.}\ \bibnamefont {Veldhorst}}, \bibinfo
  {author} {\bibfnamefont {T.~B.}\ \bibnamefont {Schr{\o}der}},\ and\ \bibinfo
  {author} {\bibfnamefont {J.~C.}\ \bibnamefont {Dyre}},\ }\bibfield  {title}
  {\bibinfo {title} {Statistical mechanics of roskilde liquids: configurational
  adiabats, specific heat contours, and density dependence of the scaling
  exponent},\ }\href@noop {} {\bibfield  {journal} {\bibinfo  {journal} {J.
  Chem. Phys.}\ }\textbf {\bibinfo {volume} {139}},\ \bibinfo {pages} {184506}
  (\bibinfo {year} {2013})}\BibitemShut {NoStop}%
\bibitem [{\citenamefont {Rosenfeld}(1977)}]{Rosenfeld1977}%
  \BibitemOpen
  \bibfield  {author} {\bibinfo {author} {\bibfnamefont {Y.}~\bibnamefont
  {Rosenfeld}},\ }\bibfield  {title} {\bibinfo {title} {Relation between the
  transport coefficients and the internal entropy of simple systems},\ }\href
  {https://doi.org/10.1103/physreva.15.2545} {\bibfield  {journal} {\bibinfo
  {journal} {Phys. Rev. A}\ }\textbf {\bibinfo {volume} {15}},\ \bibinfo
  {pages} {2545} (\bibinfo {year} {1977})}\BibitemShut {NoStop}%
\bibitem [{\citenamefont {Dyre}(2018)}]{Dyre:2018}%
  \BibitemOpen
  \bibfield  {author} {\bibinfo {author} {\bibfnamefont {J.~C.}\ \bibnamefont
  {Dyre}},\ }\bibfield  {title} {\bibinfo {title} {Isomorph theory of physical
  aging},\ }\href {https://doi.org/10.1063/1.5022999} {\bibfield  {journal}
  {\bibinfo  {journal} {J. Chem. Phys.}\ }\textbf {\bibinfo {volume} {148}},\
  \bibinfo {pages} {154502} (\bibinfo {year} {2018})}\BibitemShut {NoStop}%
\bibitem [{\citenamefont {Dyre}(2020)}]{Dyre:2020}%
  \BibitemOpen
  \bibfield  {author} {\bibinfo {author} {\bibfnamefont {J.~C.}\ \bibnamefont
  {Dyre}},\ }\bibfield  {title} {\bibinfo {title} {Isomorph theory beyond
  thermal equilibrium},\ }\href@noop {} {\bibfield  {journal} {\bibinfo
  {journal} {J. Chem. Phys.}\ }\textbf {\bibinfo {volume} {153}},\ \bibinfo
  {pages} {134502} (\bibinfo {year} {2020})}\BibitemShut {NoStop}%
\bibitem [{\citenamefont {Kob}\ and\ \citenamefont
  {Andersen}(1994)}]{Kob/Andersen:1994}%
  \BibitemOpen
  \bibfield  {author} {\bibinfo {author} {\bibfnamefont {W.}~\bibnamefont
  {Kob}}\ and\ \bibinfo {author} {\bibfnamefont {H.~C.}\ \bibnamefont
  {Andersen}},\ }\bibfield  {title} {\bibinfo {title} {{Scaling behavior in the
  $\beta$-Relaxation Regime of a Supercooled {Lennard-Jones} Mixture}},\
  }\href@noop {} {\bibfield  {journal} {\bibinfo  {journal} {Phys. Rev. Lett.}\
  }\textbf {\bibinfo {volume} {73}},\ \bibinfo {pages} {1376} (\bibinfo {year}
  {1994})}\BibitemShut {NoStop}%
\bibitem [{\citenamefont {Kob}\ and\ \citenamefont
  {Andersen}(1995{\natexlab{a}})}]{Kob/Andersen:1995a}%
  \BibitemOpen
  \bibfield  {author} {\bibinfo {author} {\bibfnamefont {W.}~\bibnamefont
  {Kob}}\ and\ \bibinfo {author} {\bibfnamefont {H.~C.}\ \bibnamefont
  {Andersen}},\ }\bibfield  {title} {\bibinfo {title} {{Testing mode-coupling
  theory for a supercooled binary Lennard-Jones mixture I: The van Hove
  correlation function}},\ }\href@noop {} {\bibfield  {journal} {\bibinfo
  {journal} {Phys. Rev. E}\ }\textbf {\bibinfo {volume} {51}},\ \bibinfo
  {pages} {4626} (\bibinfo {year} {1995}{\natexlab{a}})}\BibitemShut {NoStop}%
\bibitem [{\citenamefont {Kob}\ and\ \citenamefont
  {Andersen}(1995{\natexlab{b}})}]{Kob/Andersen:1995b}%
  \BibitemOpen
  \bibfield  {author} {\bibinfo {author} {\bibfnamefont {W.}~\bibnamefont
  {Kob}}\ and\ \bibinfo {author} {\bibfnamefont {H.~C.}\ \bibnamefont
  {Andersen}},\ }\bibfield  {title} {\bibinfo {title} {{Testing mode-coupling
  theory for a supercooled binary Lennard-Jones mixture. II. Intermediate
  scattering function and dynamic susceptibility}},\ }\href@noop {} {\bibfield
  {journal} {\bibinfo  {journal} {Phys. Rev. E}\ }\textbf {\bibinfo {volume}
  {52}},\ \bibinfo {pages} {4134} (\bibinfo {year}
  {1995}{\natexlab{b}})}\BibitemShut {NoStop}%
\bibitem [{\citenamefont {Evans}\ and\ \citenamefont
  {Morriss}(1984)}]{Evans/Morriss:1984}%
  \BibitemOpen
  \bibfield  {author} {\bibinfo {author} {\bibfnamefont {D.~J.}\ \bibnamefont
  {Evans}}\ and\ \bibinfo {author} {\bibfnamefont {G.~P.}\ \bibnamefont
  {Morriss}},\ }\bibfield  {title} {\bibinfo {title} {{Nonlinear-response
  theory for steady planar couette-flow}},\ }\href
  {https://doi.org/10.1103/PhysRevA.30.1528} {\bibfield  {journal} {\bibinfo
  {journal} {Phys.\ Rev.\ A}\ }\textbf {\bibinfo {volume} {30}},\ \bibinfo
  {pages} {1528} (\bibinfo {year} {1984})}\BibitemShut {NoStop}%
\bibitem [{\citenamefont {Ladd}(1984)}]{Ladd:1984}%
  \BibitemOpen
  \bibfield  {author} {\bibinfo {author} {\bibfnamefont {A.~J.~C.}\
  \bibnamefont {Ladd}},\ }\bibfield  {title} {\bibinfo {title} {{Equations of
  motion for non-equilibrium molecular-dynamics simulations of viscous-flow in
  molecular liquids}},\ }\href {https://doi.org/10.1080/00268978400102441}
  {\bibfield  {journal} {\bibinfo  {journal} {Mol. Phys.}\ }\textbf {\bibinfo
  {volume} {53}},\ \bibinfo {pages} {459} (\bibinfo {year} {1984})}\BibitemShut
  {NoStop}%
\bibitem [{\citenamefont {Lees/Edwards}(1972)}]{Lees/Edwards:1972}%
  \BibitemOpen
  \bibfield  {author} {\bibinfo {author} {\bibnamefont {Lees/Edwards}},\
  }\bibfield  {title} {\bibinfo {title} {The computer study of transport
  processes under extreme conditions},\ }\href@noop {} {\bibfield  {journal}
  {\bibinfo  {journal} {J. Phys. C: Solid State Phys}\ }\textbf {\bibinfo
  {volume} {5}},\ \bibinfo {pages} {1921} (\bibinfo {year} {1972})}\BibitemShut
  {NoStop}%
\bibitem [{\citenamefont {Allen}\ and\ \citenamefont
  {Tildesley}(1987)}]{Allen/Tildesley:1987}%
  \BibitemOpen
  \bibfield  {author} {\bibinfo {author} {\bibfnamefont {M.~P.}\ \bibnamefont
  {Allen}}\ and\ \bibinfo {author} {\bibfnamefont {D.~J.}\ \bibnamefont
  {Tildesley}},\ }\href@noop {} {\emph {\bibinfo {title} {Computer Simulation
  of Liquids}}}\ (\bibinfo  {publisher} {Oxford University Press},\ \bibinfo
  {year} {1987})\BibitemShut {NoStop}%
\bibitem [{\citenamefont {Bailey}\ \emph {et~al.}(2017)\citenamefont {Bailey},
  \citenamefont {Ingebrigtsen}, \citenamefont {Hansen}, \citenamefont
  {Veldhorst}, \citenamefont {B{\o}hling}, \citenamefont {Lemarchand},
  \citenamefont {Olsen}, \citenamefont {Bacher}, \citenamefont {Costigliola},
  \citenamefont {Pedersen}, \citenamefont {Larsen}, \citenamefont {Dyre},\ and\
  \citenamefont {Schr{\o}der}}]{Bailey2017}%
  \BibitemOpen
  \bibfield  {author} {\bibinfo {author} {\bibfnamefont {N.}~\bibnamefont
  {Bailey}}, \bibinfo {author} {\bibfnamefont {T.}~\bibnamefont
  {Ingebrigtsen}}, \bibinfo {author} {\bibfnamefont {J.~S.}\ \bibnamefont
  {Hansen}}, \bibinfo {author} {\bibfnamefont {A.}~\bibnamefont {Veldhorst}},
  \bibinfo {author} {\bibfnamefont {L.}~\bibnamefont {B{\o}hling}}, \bibinfo
  {author} {\bibfnamefont {C.}~\bibnamefont {Lemarchand}}, \bibinfo {author}
  {\bibfnamefont {A.}~\bibnamefont {Olsen}}, \bibinfo {author} {\bibfnamefont
  {A.}~\bibnamefont {Bacher}}, \bibinfo {author} {\bibfnamefont
  {L.}~\bibnamefont {Costigliola}}, \bibinfo {author} {\bibfnamefont
  {U.}~\bibnamefont {Pedersen}}, \bibinfo {author} {\bibfnamefont
  {H.}~\bibnamefont {Larsen}}, \bibinfo {author} {\bibfnamefont
  {J.}~\bibnamefont {Dyre}},\ and\ \bibinfo {author} {\bibfnamefont
  {T.}~\bibnamefont {Schr{\o}der}},\ }\bibfield  {title} {\bibinfo {title}
  {{RUMD}: A general purpose molecular dynamics package optimized to utilize
  {GPU} hardware down to a few thousand particles},\ }\bibfield  {journal}
  {\bibinfo  {journal} {{SciPost} Physics}\ }\textbf {\bibinfo {volume} {3}},\
  \href {https://doi.org/10.21468/scipostphys.3.6.038}
  {10.21468/scipostphys.3.6.038} (\bibinfo {year} {2017})\BibitemShut {NoStop}%
\bibitem [{\citenamefont {Separdar}\ \emph {et~al.}(2013)\citenamefont
  {Separdar}, \citenamefont {Bailey}, \citenamefont {Schr{\o}der},
  \citenamefont {Davatolhagh},\ and\ \citenamefont {Dyre}}]{Separdar2013}%
  \BibitemOpen
  \bibfield  {author} {\bibinfo {author} {\bibfnamefont {L.}~\bibnamefont
  {Separdar}}, \bibinfo {author} {\bibfnamefont {N.~P.}\ \bibnamefont
  {Bailey}}, \bibinfo {author} {\bibfnamefont {T.~B.}\ \bibnamefont
  {Schr{\o}der}}, \bibinfo {author} {\bibfnamefont {S.}~\bibnamefont
  {Davatolhagh}},\ and\ \bibinfo {author} {\bibfnamefont {J.~C.}\ \bibnamefont
  {Dyre}},\ }\bibfield  {title} {\bibinfo {title} {Isomorph invariance of
  couette shear flows simulated by the {SLLOD} equations of motion},\ }\href
  {https://doi.org/10.1063/1.4799273} {\bibfield  {journal} {\bibinfo
  {journal} {J. Chem. Phys.}\ }\textbf {\bibinfo {volume} {138}},\ \bibinfo
  {pages} {154505} (\bibinfo {year} {2013})}\BibitemShut {NoStop}%
\bibitem [{\citenamefont {Jiang}\ \emph {et~al.}(2019)\citenamefont {Jiang},
  \citenamefont {Weeks},\ and\ \citenamefont {Bailey}}]{Jiang2019}%
  \BibitemOpen
  \bibfield  {author} {\bibinfo {author} {\bibfnamefont {Y.}~\bibnamefont
  {Jiang}}, \bibinfo {author} {\bibfnamefont {E.~R.}\ \bibnamefont {Weeks}},\
  and\ \bibinfo {author} {\bibfnamefont {N.~P.}\ \bibnamefont {Bailey}},\
  }\bibfield  {title} {\bibinfo {title} {Isomorph invariance of dynamics of
  sheared glassy systems},\ }\bibfield  {journal} {\bibinfo  {journal}
  {Physical Review E}\ }\textbf {\bibinfo {volume} {100}},\ \href
  {https://doi.org/10.1103/physreve.100.053005} {10.1103/physreve.100.053005}
  (\bibinfo {year} {2019})\BibitemShut {NoStop}%
\bibitem [{\citenamefont {Schr{\o}der}(2021)}]{Schroder2021}%
  \BibitemOpen
  \bibfield  {author} {\bibinfo {author} {\bibfnamefont {T.~B.}\ \bibnamefont
  {Schr{\o}der}},\ }\href@noop {} {\bibinfo {title} {{Predicting scaling
  properties from a single fluid configuration }}} (\bibinfo {year} {2021}),\
  \Eprint {https://arxiv.org/abs/arXiv:2105.12258} {arXiv:2105.12258}
  \BibitemShut {NoStop}%
\bibitem [{\citenamefont {Toxvaerd}\ and\ \citenamefont
  {Dyre}(2011)}]{Toxvaerd2011}%
  \BibitemOpen
  \bibfield  {author} {\bibinfo {author} {\bibfnamefont {S.}~\bibnamefont
  {Toxvaerd}}\ and\ \bibinfo {author} {\bibfnamefont {J.~C.}\ \bibnamefont
  {Dyre}},\ }\bibfield  {title} {\bibinfo {title} {Communication: Shifted
  forces in molecular dynamics},\ }\href {https://doi.org/10.1063/1.3558787}
  {\bibfield  {journal} {\bibinfo  {journal} {J. Chem. Phys.}\ }\textbf
  {\bibinfo {volume} {134}},\ \bibinfo {pages} {081102} (\bibinfo {year}
  {2011})},\ \Eprint {https://arxiv.org/abs/https://doi.org/10.1063/1.3558787}
  {https://doi.org/10.1063/1.3558787} \BibitemShut {NoStop}%
\bibitem [{\citenamefont {Pedersen}\ \emph {et~al.}(2018)\citenamefont
  {Pedersen}, \citenamefont {Schr{\o}der},\ and\ \citenamefont
  {Dyre}}]{Pedersen/Schroder/Dyre:2018}%
  \BibitemOpen
  \bibfield  {author} {\bibinfo {author} {\bibfnamefont {U.~R.}\ \bibnamefont
  {Pedersen}}, \bibinfo {author} {\bibfnamefont {T.~B.}\ \bibnamefont
  {Schr{\o}der}},\ and\ \bibinfo {author} {\bibfnamefont {J.~C.}\ \bibnamefont
  {Dyre}},\ }\bibfield  {title} {\bibinfo {title} {{Phase Diagram of
  Kob-Andersen-Type Binary Lennard-Jones Mixtures}},\ }\href
  {https://doi.org/10.1103/PhysRevLett.120.165501} {\bibfield  {journal}
  {\bibinfo  {journal} {Phys. Rev. Lett.}\ }\textbf {\bibinfo {volume} {120}},\
  \bibinfo {pages} {165501} (\bibinfo {year} {2018})}\BibitemShut {NoStop}%
\bibitem [{\citenamefont {Evans}\ and\ \citenamefont
  {Morriss}(2008)}]{Evans/Morriss:2008}%
  \BibitemOpen
  \bibfield  {author} {\bibinfo {author} {\bibfnamefont {D.~J.}\ \bibnamefont
  {Evans}}\ and\ \bibinfo {author} {\bibfnamefont {G.}~\bibnamefont
  {Morriss}},\ }\href@noop {} {\emph {\bibinfo {title} {{Statistical Mechanics
  of Nonequilibrium Liquids}}}},\ \bibinfo {edition} {2nd}\ ed.\ (\bibinfo
  {publisher} {Cambridge University Press},\ \bibinfo {year}
  {2008})\BibitemShut {NoStop}%
\bibitem [{\citenamefont {Spearman}(1904)}]{Spearman:1904}%
  \BibitemOpen
  \bibfield  {author} {\bibinfo {author} {\bibfnamefont {C.}~\bibnamefont
  {Spearman}},\ }\bibfield  {title} {\bibinfo {title} {General intelligence,
  objectively determined and measured},\ }\href
  {https://doi.org/10.2307/1412107} {\bibfield  {journal} {\bibinfo  {journal}
  {Am. J. Psychol.}\ }\textbf {\bibinfo {volume} {15}},\ \bibinfo {pages} {201}
  (\bibinfo {year} {1904})}\BibitemShut {NoStop}%
\bibitem [{\citenamefont {Kendall}\ and\ \citenamefont
  {Gibbons}(1990)}]{Kendall/Gibbons:1990}%
  \BibitemOpen
  \bibfield  {author} {\bibinfo {author} {\bibfnamefont {M.~G.}\ \bibnamefont
  {Kendall}}\ and\ \bibinfo {author} {\bibfnamefont {J.~D.}\ \bibnamefont
  {Gibbons}},\ }\href@noop {} {\emph {\bibinfo {title} {Rank Correlation
  Methods}}},\ \bibinfo {edition} {5th}\ ed.\ (\bibinfo  {publisher} {Oxford
  University Press},\ \bibinfo {year} {1990})\BibitemShut {NoStop}%
\end{thebibliography}%

\end{document}